\documentclass[]{article}

\usepackage{amsmath,latexsym,amssymb,multicol,graphics}

\newcommand{\tr}{\ensuremath{\mathrm{tr}}}

\newcommand{\im}{\ensuremath{\mathrm{Im}}}

\newcommand{\cn}{\ensuremath{\mathrm{cn}}}
\newcommand{\sn}{\ensuremath{\mathrm{sn}}}
\newcommand{\dn}{\ensuremath{\mathrm{dn}}}
\newcommand{\tp}{two-point boundary value problem }
\newcommand{\tps}{two-point boundary value problems }
\newcommand{\tpp}{two-point boundary value problem. }

\newcommand{\otp}{optimal transfer problem }
\newcommand{\otps}{optimal transfer problems }
\newcommand{\otpp}{optimal transfer problem. }

\begin{document}

\title{{\Large{Optimal population transfers in a
quantum system for large transfer time}}}

\author{Symeon Grivopoulos \\[-0.1cm]
{\footnotesize \textsf{symeon@engr.ucsb.edu}}
\and Bassam Bamieh \\[-0.1cm]
{\footnotesize \textsf{bamieh@engr.ucsb.edu}}}

\maketitle

\begin{center}
\small \textsf{Department of Mechanical and Environmental Engineering} \\
\small \textsf{University of California, Santa Barbara, CA
93106-5070}
\end{center}

\begin{abstract}
Transferring the state of a quantum system to a given distribution
of populations is an important problem with applications to
Quantum Chemistry and Atomic Physics. In this work we consider
exact population transfers that minimize the $L^{2}$ norm of the
control which is typically the amplitude of an electromagnetic
field. This problem is analytically and numerically challenging.
Except for few exactly solvable cases, there is no general
understanding of the nature of optimal controls and trajectories.
We find that by examining the limit of large transfer times, we
can uncover such general properties. In particular, for transfer
times large with respect to the time scale of the free dynamics of
the quantum system, the optimal control is a sum of components,
each being a Bohr frequency sinusoid modulated by a slow
amplitude, i.e. a profile that changes considerably only on the
scale of the transfer time. Moreover, we show that the optimal
trajectory follows a ``mean'' evolution modulated by the fast free
dynamics of the system. The calculation of the ``mean'' optimal
trajectory and the slow control profiles is done via an
``averaged'' \tp which we derive and which is much easier to solve
than the one expressing the necessary conditions for optimality of
the original \otpp
\end{abstract}

\section{Introduction}
\label{Introduction}

Steering a quantum system from its initial state to a given final
state or a set of final states is one of the central problems in
the control of such systems. While transfers to specific final
states are very important for applications to Quantum Computing
and Quantum Chemistry, transfers to given final populations are
also important for many applications to Quantum Chemistry and
Atomic Physics. Optimal Control is a natural approach to transfer
problems: Frequently, one desires to optimize some aspect of the
transfer. For example, minimize the transfer time
\cite{khabrogla01a,dal02}, maximize some measure of efficiency of
the control in achieving its objective \cite{khareiluygla02} or
minimize some measure of the size of the control, for instance its
$L^2$ norm  \cite{daldah01, boschagau02, boschagauguejau02,
sheshirab93}. Moreover, the optimal control(s), singled out of all
possible controls that achieve the objective, should have
interesting properties tied to the structure of the given system.

In this work we consider exact population transfers that minimize
the $L^{2}$ norm of the control. We work in the semiclassical
approximation, where the control field influencing the quantum
system is taken to be a classical source. We also employ the
dipole moment approximation, valid for long wavelength fields,
where the spatial features of the system-control field interaction
are lumped in an interaction Hamiltonian and the amplitude of the
field becomes the control parameter. The dynamics of the system is
given then by the time-dependent Schrodinger equation (we consider
a system with one control - which is usually the case - and set
$\hbar$ to $1$):
\begin{equation} \label{simple controlled Schrodinger}
i \dot{\psi} = ( H_{0} + V\,u(t))\psi.
\end{equation}
We wish to find a $u \in \, L^{2}([0,T])$, that minimizes
\begin{equation}  \label{integral cost of the control}
\|u\|^{2}_{L^{2}([0,T])}= \int_{0}^{T}  u^{2}(t) \,dt,
\end{equation}
and drives an initial state $\psi_{0}$ of system (\ref{simple
controlled Schrodinger}) to a target population distribution
$\{|\psi_{i}(T)|^{2}=p_{i},\ i=1,\ldots,N \}$ ($N$ is the
dimension of the system and $\psi=\sum_{i}\psi_{i}e_{i}$, where
the $e_{i}$'s are the orthonormal eigenvectors of $H_{0}$). This
cost has been used extensively in the literature of optimal
control of quantum systems. It provides a measure of the energy
spent to create the controlling field and leads, as we will see,
to interesting conclusions.

Relatively little work has been done on this problem.
Analytically, it is a hard problem and explicit solutions are
known only in few cases: In \cite{daldah01,boschagauguejau02} this
is done for transfers between eigenstates in a two-dimensional
system with two controls and in
\cite{boschagau02,boschagauguejau02} it is done for transfers
between eigenstates of a three-dimensional system with four
controls. In fact, both cases mentioned above are instances of a
certain algebraic structure being present (the so-called $K+P$
structure) \cite{boschagau02,jur01}, that allows one to find
analytic expressions for the control and the state in terms of the
unknown initial costate but the analytic determination of this
unknown initial costate (and hence the complete solution of the
problem) is possible only for systems of small dimensionality
because one needs to analytically compute matrix exponentials and
solve transcendental equations. In any event, this structure is
special and will not be present in general. In another case
\cite{daldah01} (for a two-state system with one control), it is
possible to find an analytic expression for the control (in terms
of unknown constants) but not for the state and so the problem has
to be solved numerically from that point on. For higher
dimensional systems, no general properties of the optimal control
and state trajectory are known.

Numerically, the \tp that expresses the necessary conditions of
optimality becomes increasingly harder to solve as the dimension
of the system or the transfer time grows. Dimension growth
dramatically increases the computational cost of the numerical
solution. More relevant to our work is the issue of large transfer
time. In many typical applications, the transfer time may be a few
orders of magnitude larger than the time scale of the free
evolution of the system. This may be necessary for the transfer to
be possible or for the amplitude of the control to be within
experimentally feasible limits. The problem here is the presence
of two time scales in the solution, i.e. stiffness: There is the
fast time scale of the free dynamics of the quantum system and the
slow time scale of the transition (which is of the order of $T$).
This creates the need for a very detailed numerical solution in
order to guarantee good solution accuracy resulting in large
computational times. In applications one usually considers
relaxations of this \otp where one abandons the requirement for
\emph{exact} transfers and tries to minimize a combination of the
integral cost (\ref{integral cost of the control}) and a measure
of distance from the desired final state or population
distribution such as
\begin{equation}\label{relaxed cost functional}
a \int_{0}^{T} u^{2}(t) \,dt \, +
\sum_{i=1}^{N}\big(|\psi_{i}(T)|^{2}-p_{i} \big)^2,
\end{equation}
($a \geq 0$). An advantage of \otps like these over the one we are
considering in this work is that they lead to \tps with separated
boundary conditions and are amenable to iterative solution
methods, see for example \cite{peidahrab87,zhurab98,gribam02}.
Yet, the above remarks hold for these as well. We shall comment on
the connections between this work and the later type of \otps in
section \ref{Conclusion}.

We found that the study of optimal population transfers for large
transfer times offers some insight into the nature of
optimal control and state trajectory as well as computational
advantages in the numerical solution of the problem. The main
conclusions of our work are the following:
\begin{enumerate}
\item For generic population transfers and large enough transfer
times, the optimal control has the following, physically plausible
form: It is a sum of sinusoids with frequencies equal to the Bohr
frequencies of the quantum system multiplied by slowly varying
profiles, that is functions of $\frac{t}{T}$:
\begin{eqnarray*}
u_{opt}(t)&=&\frac{i}{T} \tr \big( e^{i H_{0}t}\, V\,e^{-i H_{0}t} L(\frac{t}{T}) \big) + O(\frac{1}{T^{2}})  \\
&=&\frac{i}{T} \sum_{k \neq l}V_{kl}\, e^{i\omega_{kl} t}
L_{lk}(\frac{t}{T}) + O(\frac{1}{T^{2}}).
\end{eqnarray*}
$L$ is an anti-Hermitian matrix with zeros on the diagonal, whose
entries are the profiles. This form is explicitly verified in all
analytically solvable cases (where, because of the structure of the systems, there are no $O(\frac{1}{T^2})$ corrections) and it is
in fact observed in numerical solutions.
\item Again for generic population transfers and large enough
transfer times, the optimal trajectories follow a slow ``mean''
evolution upon which small deviations are imposed. The slow mean
evolution and the slow control profiles can be calculated by
solving an ``averaged'' \tp (the term will be explained in section
\ref{Optimal population transfers for an averaged system}) in the
fixed interval $[0,1]$, irrespective of how large the transfer
time $T$ is. The small deviations of the optimal trajectories from
their mean evolution are due to the free dynamics. Quantitatively,
\[ \psi(t)=e^{-i H_{0}t}\,\bar{\psi}(\frac{t}{T}) + O(\frac{1}{T}), \]
where $\bar{\psi}$ denotes the ``mean trajectory'' and the
oscillatory $e^{-i H_{0}t}$ term is responsible for the deviations
from the mean evolution. There are also additional corrections, of
higher order in a $\frac{1}{T}$ expansion.
\end{enumerate}
Although these results hold for large transfer times, one may use
a solution of the optimal control problem obtained this way for a
large transfer time $T$ as the first step in a continuation method
solution of the original optimization problem where the
continuation parameter is the transfer time. The point is that the
large transfer time limit both reveals the structure of the
controls and serves as a good starting point for the solution of
the problem: Indeed, the associated ``averaged'' \tp in $[0,1]$ is
much easier to solve numerically than the original problem.

Our presentation is organized as follows: In section \ref{Optimal
population transfers}, we set up the optimal population transfer
problem for a finite dimensional quantum system, review approaches
to its solution and comment on the difficulties associated with
these approaches. In section \ref{Optimal population transfers for
an averaged system}, we consider the quantum system with a certain
class of controls and show how, for large transfer times $T$, it
is approximated to first order in a $\frac{1}{T}$ perturbation
expansion by an ``averaged'' control system. Then, we set up an
optimal population transfer problem for this averaged system.
Section \ref{Main Results} contains our main result, namely that,
for large transfer time, the solution to the original optimal
control problem is approximated by the solution of the optimal
population transfer problem for the averaged system. This implies
that the optimal control for the original problem belongs to the
class of controls used to transform the original system to the
averaged one and this provides a useful characterization of it.
The proof of this result is contained in section \ref{Proof of
Main Results}. We demonstrate our approach with some examples in
section \ref{Examples}. Section \ref{Conclusion} concludes.

\section{Optimal population transfers}
\label{Optimal population transfers}

In this section we derive the necessary conditions of optimality
for the optimal population transfer problem described in section
\ref{Introduction}. We consider a quantum system influenced by an
external, classical source (for example the electric field of a
laser). For simplicity, we consider a system with one control
(which is usually the case) and set $\hbar$ to $1$ in
Schrodinger's equation:
\[ i \dot{\psi} = ( H_{0} + V\,u(t))\psi. \]
Nevertheless, everything we do in this and the following sections
can readily be generalized to systems with more controls.

Before we consider the optimal control problem, we want to address
the controllability question for this system. The system
(\ref{simple controlled Schrodinger}) is controllable if, for
every pair of initial and target states $(\psi_{0},\, \psi_{d})$
there exists a transfer time $T$ and a measurable $u(t), \ t \in
[0,T]$, such that the solution of (\ref{simple controlled
Schrodinger}) with $u(t)$ and $\psi(0) = e^{i
\varphi_0}\,\psi_{0}$ results into $\psi(T) = e^{i \varphi_d}\,
\psi_{d}$, for some $\varphi_0, \,  \varphi_d \, \in S^{1}$. (In
Quantum Mechanics all states are defined modulo a total phase.
From (\ref{simple controlled Schrodinger}) it is obvious that one
may always set $\varphi_0=0$, for example.) Sufficiency conditions
for controllability \cite{ramsaldahrabpei95,alt02a,turrab01} are
based on the classical results \cite{jursus72,bro72,bro73}. In
this work, we would like to assume a somewhat strong form of
controllability assumption as given, for example, in
\cite{alt02a}. To state it, we need a few simple notions: A
quantum system whose energy levels (eigenvalues of $H_0$) are all
different from each other is called non-degenerate. Moreover, a
system such that no two Bohr frequencies (differences of energy
levels) are the same, is said to have no degenerate transitions.
The connectivity graph ${\mathcal{G}}_{M}$ of a Hermitian matrix
$M=(M_{ij})$ is defined as the pair $({\mathcal{N}}_{M},
{\mathcal{C}}_{M})$, where ${\mathcal{N}}_{M} = \{ 1,\ldots,N \}$
is the set of nodes and ${\mathcal{C}}_{M} = \{ (i,j), \ i>j, \
|M_{ij}\neq 0 \}$ is the set of edges joining the nodes. The graph
${\mathcal{G}}_{M}$ is called connected when there exists a path
joining every two nodes. \vspace*{.7em} \\
{\bf Controllability assumption:} The system (\ref{simple
controlled Schrodinger})  is non-degenerate, has no degenerate
transitions and the graph of $V$ is connected. \vspace*{.7em} \\
Then \cite{alt02a}, (\ref{simple controlled Schrodinger}) is
controllable. As a matter of fact, this is the generic situation
for controllability of (\ref{simple controlled Schrodinger}).

We now proceed with the set up of the optimal control problem.
We wish to find a $u \in \, L^{2}([0,T])$, that minimizes
\[  \|u\|^{2}_{L^{2}([0,T])}= \int_{0}^{T}  u^{2}(t) \,dt, \]
and drives an initial state $\psi_{0}$ of system (\ref{simple
controlled Schrodinger}) to a target population distribution
$\{|\psi_{i}(T)|^{2}=p_{i},\ i=1,\ldots,N \}$ ($N$ is the
dimension of the system). We will refer to this as \otp (I). The
Maximum Principle of optimal control \cite{jur97,sagwhi77}
provides necessary conditions for optimality in terms of the
Hamiltonian function
\[ H(\psi,\lambda,u)=\frac{1}{2} u^{2}-i \lambda^{*}( H_{0} + V\,u)\psi
+ i \psi^{*}( H_{0} + V\,u)\lambda, \]
where $\lambda \in \mathbb{C}^{n}$ is the co-state vector. It is
shown in Appendix A that solutions of the optimal control problem
satisfy, $\lambda^{*}(t)\,\psi(t)=0$.

The optimal control $u$ and the corresponding state and co-state
satisfy the equations
\[ \dot{\psi} = \frac{\partial H}{\partial \lambda^{*}},\ \
\dot{\lambda}= -\frac{\partial H}{\partial \psi^{*}}\ \ \mathrm{and} \
\ \frac{\partial H}{\partial u}=0, \]
which, given the expression for $H$, have the form
\begin{eqnarray}
i \dot{\psi} &=& (H_{0} + V\,u)\psi,\label{oc state} \\
i \dot{\lambda} &=& (H_{0} + V\,u)\lambda ,\label{oc costate} \\
u &=& i\,(\lambda^{*}V\psi - \psi^{*}V\lambda). \label{oc control}
\end{eqnarray}
For convenience, we rewrite equations (\ref{oc state}) and
(\ref{oc costate}) using $u$ from (\ref{oc control}):
\begin{eqnarray}
i \dot{\psi} &=& H_{0}\psi +  i\,(\lambda^{*}V\psi - \psi^{*}V\lambda)\, V\psi,
\label{oc nl state} \\
i \dot{\lambda} &=& H_{0}\lambda +  i\,(\lambda^{*}V\psi - \psi^{*}V\lambda)\, V\lambda.
\label{oc nl costate}
\end{eqnarray}
To these equations one must append the boundary conditions
\begin{equation} \label{oc boundary conditions}
\psi(0)=\psi_{0}, \ \ \  |\psi_{i}(T)|^{2}=p_{i}, \ \ \
\im(\psi^{*}_{i}(T)\,\lambda_{i}(T))=0,
\end{equation}
$\forall \ i=1,\ldots,N$. The last of these equations  are the
transversality conditions at the endpoint. Their proof is also
given in Appendix A. We will refer to the \tp comprised of
equations (\ref{oc nl state}) - (\ref {oc boundary conditions}) as
\tp (I).

We briefly discuss now methods for its solution. Analytically, we
may proceed as follows: Define the traceless, anti-Hermitian
matrix $\Lambda$ by $ \Lambda :=
\psi\,\lambda^{*}-\lambda\,\psi^{*} $ and using (\ref{oc state})
and (\ref{oc costate}) we can show that
\begin{equation}\label{su(n) costate eq}
\dot{\Lambda}=-i\,[H_{0} + V\,u, \Lambda] = [-i H_{0} +
\tr(V\Lambda)\,V ,\Lambda]
\end{equation}
where we substituted
\[ u = i \,(\lambda^{*}V\psi - \psi^{*}V\lambda) =i \,\tr(V\Lambda), \]
from (\ref{oc control}). If we solve (\ref{su(n) costate eq})
analytically, we obtain an expression for the control function
$u(t)$ in terms of unknown constants (these unknown constants are
matrix elements of
$\Lambda(0)=\psi(0)\,\lambda^{*}(0)-\lambda(0)\,\psi^{*}(0)$ and
$\lambda(0)$ is what we are after). Then, one may attempt to solve
(\ref{oc state}) and (\ref{oc costate}) using this expression for
$u$ and determine these constants by satisfying the boundary
conditions (\ref{oc boundary conditions}). In practice this
program seems impossible to carry out in its entirety except for
the few special cases mentioned in the introduction.
\footnote {In all these works, and in fact in most of the
literature of Geometric Control Theory, authors look at the
``lifting'' of the system $i\hbar\,\dot{\psi}=(\,
H_{0}+\sum_{\alpha} V_{\alpha}u_{\alpha}(t)\,)\,\psi$ to $SU(N)$,
$i\hbar\,\dot{U}=(\, H_{0}+\sum_{\alpha}
V_{\alpha}u_{\alpha}(t)\,)\,U$ where the state transition matrix
$U(t)$ is defined by $\psi(t)=U(t)\,\psi(0)$. It turns out that
the formulation of the optimal control problem in the two setups
is similar and in fact, equation (\ref{su(n) costate eq}) in that
context is the co-state equation ``pulled-back'' on the cotangent
space at the identity of $SU(N)$, see \cite{bro73,jur97}.}
Numerically, one may attempt to solve the \tp (\ref{oc nl state})
- (\ref{oc boundary conditions}) using, for example, some shooting
or finite-difference method. This works for systems of small
dimensionality and small transfer times but becomes increasingly
harder as the system dimension grows and as larger transfer times
are required for the transfer to be possible.  As mentioned in the
introduction, in many typical applications the transfer time has
to be a few orders of magnitude larger than the time scale of the
free dynamics of the system in order to achieve the desired
transfer. Because the free dynamics of the system is oscillatory,
the first terms on the right sides of equations (\ref{oc nl
state}) and (\ref{oc nl costate}) create small oscillations of the
populations around their ``mean'' evolution towards their final
values. This creates the need for a very detailed numerical
solution in order to guarantee good solution accuracy. Examples of
this can be seen in the graphs of section \ref{Examples}. Besides
the usual unfavorable scaling of the solution efficiency with
dimension (an issue which we do not address), this is the main
source of difficulty of the problem. Note that these remarks are
quite general and independent of the specific numerical methods
used to solve the \tpp We will see in the following how this
problem can be overcome.

Before we leave this section we would like to point out that our
discussion so far, as well as in the following, will only concern
\emph{normal} extrema of the optimization problem (I).
\emph{Abnormal} extrema \cite{boscha03,jur97} will not be
considered. The reason is that the form of the abnormal extrema
does not depend on the exact cost used in the minimization
problem, and thus, the same abnormal extremum can be a local or
even a global minimizer to many different cost functionals. In
this sense, abnormal minimizers are not particular to one optimal
control problem and don't reflect its particular structure. So, in
this work, all the discussion and results concern normal minima of
problem (I) (as well as problem (II) to be defined in the next
section).

\section{Optimal population transfers for an averaged system}
\label{Optimal population transfers for an averaged system}

In this section, we introduce a special form for the control in
equation (\ref{simple controlled Schrodinger}), a sum of sinusoids
with frequencies equal to the Bohr frequencies of the quantum
system multiplied by slowly varying profiles, that is functions of
$\frac{t}{T}$. We then proceed to ``average out''  the dynamics in
the time scale of the free evolution of the system (this time
scale is set by the Bohr frequencies), which is fast compared to
the transfer time $T$. This leaves us with an ``averaged''
control system whose evolution approximates that of the original
under the special form of the control introduced. The motivation
for this lies in the following: We set up a corresponding \otp for
the averaged system, whose cost approximates the cost
(\ref{integral cost of the control}). We will show in section
\ref{Proof of Main Results} how solutions to this \otp approximate
solutions to our original \otp (I), to first order in an
$O(\frac{1}{T})$ expansion, proving the results described in the
introduction. We begin with the change of variable
\begin{equation}  \label{ip}
x = e^{i H_{0}t}\psi
\end{equation}
in (\ref{simple controlled Schrodinger}). In Physics, this is
referred to as ``transforming to the Interaction Picture''. The
time evolution of the new variable $x$ is due entirely to the
control, because the free evolution has been accounted for. In
terms of the new variable, (\ref{simple controlled Schrodinger})
becomes
\begin{equation} \label{simple system ip}
i \dot{x} = u\, F(t)x,
\end{equation}
where,
\begin{equation} \label{definition of F(t)}
F(t) :=  e^{i H_{0}t}\, V\,e^{-i H_{0}t}.
\end{equation}
Note the appearance of the Bohr frequencies in the matrix elements
of F,
\[ F_{ij}(t)=V_{ij}e^{i (E_{i}-E_{j})t}=V_{ij}e^{i \omega_{ij} t}. \]
We adopt the following form for the control $u(t)$:
\begin{equation}
\label{control in form of sum of Bohr frequency sinusoids time slow envelopes}
u(t)= \varepsilon \big( u_0(\varepsilon t) + \sum_{i \neq j}^{N} e^{i \omega_{ij}t} \, u_{ji}(\varepsilon t) \big) ,
\end{equation}
where $u_{ji}^{*}=u_{ij}$ and $u_0$ is real, so that $u$ is real.
$u_{ji}$ is a complex  ``envelope'' that multiplies a sinusoid
with frequency equal to the Bohr frequency for the transition i to
j. The value of $\varepsilon$ will be given shortly. We introduce
u(t) from (\ref{control in form of sum of Bohr frequency sinusoids
time slow envelopes}) in (\ref{simple system ip}) and rewrite
(\ref{simple system ip}) in component form:
\begin{equation} \label{simple system ip control in envelope form}
i \dot{x}_i = \varepsilon \big(\, u_0(\varepsilon t) + \sum_{k \neq l}
e^{i \omega_{kl}t} \, u_{lk}(\varepsilon t) \, \big) \,
\sum_j e^{i \omega_{ij} t} V_{ij} x_j.
\end{equation}

We approximate (\ref{simple system ip control in envelope
form}) for small $\varepsilon$ using averaging. In averaging, one
considers equations of the form
\begin{equation}\label{general equation to be averaged}
\dot{w}=\varepsilon f(w,t,\varepsilon),
\end{equation}
where f must be a bounded $C^{2}$ function of its arguments with
bounded derivatives up to 2nd order such that the limit
\[ f_{av}(w) := \lim_{\tau\rightarrow\infty}\frac{1}{\tau}
\int_{t}^{t+\tau}f(t',w,0)\, dt' \]
exists. A standard averaging theorem (see Chapter 8 of
\cite{kha96}) guarantees that, for sufficiently small
$\varepsilon$, the solution of \[ \dot{\bar{w}}=\varepsilon
f_{av}(\bar{w}) \] with an initial condition $O(\varepsilon)$
close to the initial condition of (\ref{general equation to be
averaged}) (i.e. $\bar{w}(0)-w(0)=O(\varepsilon)$) is
$O(\varepsilon)$ close to that of (\ref{general equation to be
averaged}) for a time interval of length
$O(\frac{1}{\varepsilon})$. Equation (\ref{simple system ip
control in envelope form}) involves two time scales, 1 and
$\frac{1}{\varepsilon}$ but we want to average only over the time
scale 1 dynamics. This is achieved as follows: Consider,
instead of (\ref{general equation to be averaged}) the following
equation:
\begin{equation} \label{equation with two time scales to be averaged}
\dot{w}=\varepsilon f(w,t,\varepsilon t,\varepsilon).
\end{equation}
Define $w_0=\varepsilon t$ and
substitute $w_0$ for $\varepsilon t$ in (\ref{equation with two
time scales to be averaged}). Then, consider the system
\begin{eqnarray*}
\dot{w} &=& \varepsilon f(w,t,w_0,\varepsilon), \\
\dot{w_0} &=& \varepsilon
\end{eqnarray*}
and apply averaging to it. The resulting averaged form of
(\ref{equation with two time scales to be averaged}) is now
\[ \dot{\bar{w}}=\varepsilon f_{av}(\bar{w},\varepsilon t), \]
where
\[ f_{av}(w,w_0) := \lim_{\tau\rightarrow\infty}\frac{1}{\tau}
\int_{t}^{t+\tau}f(w,t',w_0,0)\, dt'. \]

We now apply this to equation (\ref{simple system ip control in
envelope form}). Since the time average of $e^{i \omega t}$ is
zero for $\omega \neq 0$ and 1 for $\omega =0$, only terms with no
time dependence will contribute to the averaged equation. Letting
$\bar{x}$ be the averaged $x$, the averaged form of (\ref{simple
system ip control in envelope form}) is
\begin{equation} \label{averaged simple system}
i \dot{\bar{x}}_i = \varepsilon \big( V_{ii}\, u_0(\varepsilon t)\, \bar{x}_i
+ \sum_{j \neq i} V_{ij}\, u_{ij}(\varepsilon t)  \, \bar{x}_j \big).
\end{equation}
Taking $\varepsilon=\frac{1}{T}$ and rescaling time to
$s=\varepsilon t =\frac{t}{T}$, (\ref{averaged simple system})
becomes
\begin{equation} \label{averaged simple system in [0,1]}
i \frac{d\bar{x}_i}{ds} = V_{ii}\, u_0(s)\, \bar{x}_i
 + \sum_{j \neq i} V_{ij}\, u_{ij}(s) \, \bar{x}_j,
\end{equation}
or, in vector form,
\[ i \frac{d\bar{x}}{ds} = \tilde{V}[u_0,u_{ij}]\, \bar{x}, \]
where
\[  \tilde{V}[u_0,u_{ij}] = \left(
\begin{array}{ccc}
V_{11}u_0 & V_{12}u_{12} &  \cdots \\
V_{21}u_{21}  & V_{22}u_0 & \cdots  \\
\vdots  & \vdots & \ddots
\end{array}
\right) \, \bar{x} = \left(
\begin{array}{ccc}
V_{11}u_0 & V_{12}u_{12} &  \cdots \\
V_{12}^{*}u_{12}^{*}  & V_{22}u_0 & \cdots  \\
\vdots  & \vdots & \ddots
\end{array}
\right) \, \bar{x}. \]
By construction, every solution to (\ref{averaged simple system in
[0,1]}) with controls $u_0(s)$ and $u_{ij}(s)$ and initial state
$\bar{x}(0)=\psi(0)$, provides a solution to (\ref{simple
controlled Schrodinger}) with $u(t)$ given by (\ref{control in
form of sum of Bohr frequency sinusoids time slow envelopes}) and
initial condition $\psi(0)$, correct up to $O(\varepsilon)$ terms
for a time interval of size $\frac{1}{\varepsilon}=T$. It is shown
in Appendix B that system (\ref{averaged simple system in [0,1]})
is controllable on account of the controllability assumption on
the original system.

Since our goal is to relate optimal transfers of the original
system to optimal transfers of the averaged one, we must find an
objective for the averaged system that approximates
$\|u\|^{2}_{L^{2}([0,T])}$. So, we compute
$\|u\|^{2}_{L^{2}([0,T])}$ for $u(t)$ given by (\ref{control in
form of sum of Bohr frequency sinusoids time slow envelopes}),
with $\varepsilon=\frac{1}{T}\,$:
\begin{eqnarray*}
\int_{0}^{T}  u^{2}(t) \,dt &=&\frac{1}{T^2} \, \int_{0}^{T}
\big( u_0(\frac{t}{T}) +
\sum_{i \neq j} e^{i \omega_{ij}t} \, u_{ji}(\frac{t}{T}) \big)^{2} \, dt \\
&=& \frac{1}{T^2}  \int_{0}^{T} \bigg\{ u_0^{2}(\frac{t}{T}) +
2\sum_{i \neq j} e^{i \omega_{ij}t} \, u_{ji}(\frac{t}{T})\, u_0(\frac{t}{T})  \\
&+&\sum_{i \neq j}\sum_{k \neq l} e^{i (\omega_{ij} + \omega_{kl})t} \,
u_{ji}(\frac{t}{T}) \, u_{lk}(\frac{t}{T}) \bigg\} \, dt 
\end{eqnarray*}
\begin{eqnarray*}
&=& \frac{1}{T} \int_{0}^{1} \big\{ u_0^{2}(s) +
2\sum_{i \neq j} e^{i \omega_{ij}Ts} \, u_{ji}(s)\, u_0(s)  \\
&+&\sum_{i \neq j}\sum_{k \neq l}
e^{i (\omega_{ij} + \omega_{kl})Ts} \, u_{ji}(s) \, u_{lk}(s)\big\} \, ds \\
&=& \frac{1}{T}  \int_{0}^{1} \big[ \sum_{i \neq j}
u_{ij}(s) \, u_{ji}(s) + u_0^{2}(s) \big] \, ds + \frac{1}{T^2}\, B(T),
\end{eqnarray*}
where $B(T)$ represents terms bounded in $T$.
This last line is the result of separating the integrals into two
kinds, these without exponentials, which are explicitly retained,
and those with, which can easily be seen to scale like
$\frac{1}{T}B(T)$ after a partial integration: Indeed, for any
differentiable $f$,
\begin{eqnarray*}
\int_{0}^{1}e^{i\omega Ts} \, f(s) \, ds &=& \frac{1}{i \omega T}
\int_{0}^{1} (e^{i\omega Ts})' \, f(s) \, ds  \\
&=&\frac{1}{i \omega T} \big\{e^{i\omega T}\, f(1) - f(0)
-\int_{0}^{1} e^{i\omega Ts} \, f'(s) \, ds \big\}.
\end{eqnarray*}
So, we pose the following optimal control problem for the averaged
system: Find controls $u_0(s)$, $u_{ij}(s), \ s \in [0,1]$, that
minimize
\begin{equation}  \label{objective in averaged system}
\int_{0}^{1} \big[ \sum_{i \neq j}
u_{ij}(s) \, u_{ji}(s) + u_0^{2}(s) \big] \, ds =
\int_{0}^{1} \big[ \sum_{i \neq j}
|u_{ij}(s)|^{2}  + u_0^{2}(s) \big] \, ds
\end{equation}
and drive an initial state $\bar{x}(0)=\psi_{0}$ of system
(\ref{averaged simple system in [0,1]}) to a target population
distribution $\{|\bar{x}_{i}(1)|^{2}=p_{i},\ i=1,\ldots,N \}$. We
will refer to this as \otp (II).

The necessary conditions for optimality are derived  from the
Hamiltonian function
\begin{eqnarray*}
&&H(\bar{x}_i,\bar{z}_i,u_{ij})= \frac{1}{2} u_0^{2} +
\frac{1}{2} \sum_{i \neq j}  |u_{ij}|^{2}
-i \bar{z}^{*}\tilde{V}[u_0,u_{ij}]\,\bar{x} + i\bar{x}^{*}\tilde{V}[u_0,u_{ij}]\,\bar{z} \\
&=& \frac{1}{2} u_0^{2} +
\sum_{i \neq j} \frac{1}{2} u_{ij} u_{ji}
-i V_{ji}u_{ji} (\bar{x}_{i}\bar{z}_{j}^{*}- \bar{z}_{i}\bar{x}_{j}^{*})
-i u_0 \sum_i V_{ii} (\bar{x}_{i}\bar{z}_{i}^{*}- \bar{z}_{i}\bar{x}_{i}^{*})
\end{eqnarray*}
and have the form
\begin{eqnarray}
i \frac{d\bar{x}_i}{ds} &=&  V_{ii}\, u_0 \, \bar{x}_i +
\sum_{j \neq i} V_{ij} u_{ij} \, \bar{x}_j \ \ \ \
(i \frac{d\bar{x}}{ds} = \tilde{V}[u_0,u_{ij}]\, \bar{x}),
\label{averaged oc state} \\
i \frac{d\bar{z}_i}{ds} &=&  V_{ii}\, u_0 \, \bar{z}_i +
\sum_{j \neq i} V_{ij} u_{ij} \, \bar{z}_j \ \ \ \
(i \frac{d\bar{z}}{ds} = \tilde{V}[u_0,u_{ij}]\, \bar{z}),
\label{averaged oc costate} \\
u_{ij}&=& i V_{ji}\, (x_i z_j^{*} - z_i x_j^{*}),  \ \ \
u_0= i \sum_i V_{ii} (\bar{x}_{i}\bar{z}_{i}^{*}- \bar{z}_{i}\bar{x}_{i}^{*}).
\label{averaged oc control}
\end{eqnarray}
We rewrite equations (\ref{averaged oc state}) and
(\ref{averaged oc costate}) using $u_0$ and $u_{ij}$ from
(\ref{averaged oc control}):
\begin{eqnarray}
\frac{d\bar{x}_{i}}{ds}&=&   \sum_{j \neq i}|V_{ij}|^{2}
(\bar{x}_{i}\bar{z}_{j}^{*} - \bar{z}_{i}\bar{x}_{j}^{*})\, \bar{x}_{j}
+ \sum_{k}V_{kk} (\bar{x}_{k}\bar{z}_{k}^{*} - \bar{z}_{k}\bar{x}_{k}^{*})\, V_{ii} \bar{x}_{i},
\label{averaged nl oc state} \\
\frac{d\bar{z}_{i}}{ds}&=&   \sum_{j \neq i}|V_{ij}|^{2} (\bar{x}_{i}\bar{z}_{j}^{* }- \bar{z}_{i}\bar{x}_{j}^{*})\, \bar{z}_{j}
+ \sum_{k}V_{kk} (\bar{x}_{k}\bar{z}_{k}^{*} - \bar{z}_{k}\bar{x}_{k}^{*})\, V_{ii} \bar{z}_{i}.
\label{averaged nl oc costate}
\end{eqnarray}
The corresponding boundary conditions are given by
\begin{equation} \label{averaged oc boundary conditions}
\bar{x}(0)=\psi_{0}, \ \ \  |\bar{x}_{i}(1)|^{2}=p_{i}, \ \ \
\im(\bar{x}^{*}_{i}(1)\,\bar{z}_{i}(1))=0.
\end{equation}
We will refer to the \tp comprised of equations (\ref{averaged nl
oc state}) - (\ref{averaged oc boundary conditions}) as \tp (II).
From (\ref{averaged oc control}), we see that all $u_{ij}$ with
$V_{ij}=0$ are identically zero, as they should. We can also show
that $u_0=0$ and simplify the right sides of equations
(\ref{averaged nl oc state}) and (\ref{averaged nl oc costate}):
In a fashion similar to the construction of $\Lambda$ in section
\ref{Optimal population transfers}, we define $L$ by $L :=
\bar{x}\bar{z}^{*}-\bar{z}\bar{x}^{*}$. We also define another
anti-Hermitian matrix $K=K(\bar{x},\bar{z})=K(L)$ by,
\begin{eqnarray}
K_{ij}&:=& |V_{ij}|^{2}L_{ij}=|V_{ij}|^{2}(\bar{x}_{i}\bar{z}_{j}^{*}- \bar{z}_{i}\bar{x}_{j}^{*}), \ \ i \neq j, \label{def of K_ij}  \\
K_{ii}&:=& V_{ii} \, \sum_{k}V_{kk}L_{kk}=V_{ii} \, \sum_{k}
V_{kk} (\bar{x}_{k}\bar{z}_{k}^{*} - \bar{z}_{k}\bar{x}_{k}^{*}).  
\label{def of K_ii}
\end{eqnarray}
$K$ is the analog of $-i H_{0} +\tr(V\Lambda)\,V$ in section
(\ref{Optimal population transfers}) and is a linear function of
$L$. With this definition, (\ref{averaged nl oc state}) and (\ref{averaged nl oc costate}) read simply as
\begin{eqnarray}
\frac{d\bar{x}}{ds}& =&K(L)\bar{x}, \label{averaged oc state2}   \\
\frac{d\bar{z}}{ds}& =&K(L)\bar{z}. \label{averaged oc costate2} 
\end{eqnarray}
It is easy to see that $L$ satisfies the differential equation
\begin{equation}
\label{averaged su(n) costate eq}
\frac{dL}{ds}=[K(L),L].
\end{equation}
The $ii$-th component of this equation reads
\begin{eqnarray*}
\frac{dL_{ii}}{ds}&=&\sum_j K_{ij} L_{ji} - L_{ij} K_{ji} \\
 &=&K_{ii}L_{ii} + \sum_{j \neq i} K_{ij} L_{ji} - L_{ii}K_{ii}
 - \sum_{j \neq i} L_{ij} K_{ji} \\
&=& \sum_{j \neq i} |V_{ij}|^{2} ( L_{ij} L_{ji} - L_{ij} L_{ji}) =0,
\end{eqnarray*}
and so,
\begin{equation}
L_{ii}(s)=L_{ii}(1)=-2i\, \im(\bar{x}^{*}_{i}(1)\,\bar{z}_{i}(1))=0.
\label{L_{ii}=0}
\end{equation}
This shows that  $K_{ii}(s)=L_{ii}(s)=0$ and $u_0(s)=0$ and so, equations (\ref{averaged nl oc state}) and (\ref{averaged nl oc costate}) simplify to
\begin{eqnarray}
\frac{d\bar{x}_{i}}{ds}&=& \sum_{j \neq i}|V_{ij}|^{2}
(\bar{x}_{i}\bar{z}_{j}^{*}- \bar{z}_{i}\bar{x}_{j}^{*})\, \bar{x}_{j},
\label{simple averaged nl oc state} \\
\frac{d\bar{z}_{i}}{ds}&=&\sum_{j \neq i}|V_{ij}|^{2} (\bar{x}_{i}\bar{z}_{j}^{*}- \bar{z}_{i}\bar{x}_{j}^{*})\, \bar{z}_{j}.
\label{simple averaged nl oc costate}
\end{eqnarray}
We will refer to this set of equations (along with (\ref{averaged
oc boundary conditions})\,) as \tp (II) as well.

\section{Main Results}
\label{Main Results}

In this section, we make the connection between solutions of the
\tps (II) and (I) in the large $T$ limit.

\vspace*{.5 em} \noindent {\bf{Theorem 1:}} Let $(\bar{x}(s),\,
\bar{z}(s))$ be a solution of \tp (II) over $[0,1]$. Define
$\psi(t),\, \lambda(t) \ \mathrm{and} \ u(t), \ t \in [0,T]$, by
\begin{eqnarray}
\psi(t)&=& e^{-i H_{0}t}\,\bar{x}(\frac{t}{T}),
\label{state approx} \\
\lambda(t)&=& \frac{1}{T}\, e^{-i H_{0}t}\,\bar{z}(\frac{t}{T}),  \label{costate approx}\\
u(t)&=&\frac{i}{T}\, \tr \big( e^{i H_{0}t}\, V\,e^{-i H_{0}t} L(\frac{t}{T}) \big) =\frac{i}{T} \sum_{kl}V_{kl}\, e^{i\omega_{kl} t} L_{lk}(\frac{t}{T}).  \label{control approx}
\end{eqnarray}
Then, for large enough $T$, $\psi(t)$ satisfies the necessary
conditions of optimality (\ref{oc state}) - (\ref{oc
boundary conditions}), up to terms of order $O(\frac{1}{T})$ and
$\lambda(t)$ and $u(t)$ up to terms of order $O(\frac{1}{T^2})$.

\vspace*{.5 em} Thus, solutions of the \tp (II) provide
{\emph{approximate}} solutions to the \tp (I) for large transfer
times $T$. A natural question to ask then is, whether these
approximate solutions to \tp (I) are in fact {\emph{approximations
to solutions}} of (I), in the large $T$ limit. This is answered
positively by the following theorem:

\vspace*{.5 em} \noindent {\bf{Theorem 2:}} Let $\psi_0$ and $\{
p_i \}_{i=1,\ldots,N}$ be an initial state and a target population
of system (\ref{simple controlled Schrodinger}), respectively, and
let $(\bar{x}(s),\, \bar{z}(s))$ be a solution of \tp (II) over
$[0,1]$. Then, for almost all pairs $(\psi_0,\{ p_i \})$, a
solution of \tp (I) exists, for large enough $T$, of the form
\begin{eqnarray}
\psi(t)&=&e^{-i H_{0}t}\,\bar{x}(\frac{t}{T}) + O(\frac{1}{T}),
\label{state approx 2} \\
\lambda(t)&=& \frac{1}{T} e^{-i H_{0}t}\,\bar{z}(\frac{t}{T})+ O(\frac{1}{T^{2}}),  \label{costate approx 2}
\end{eqnarray}
The corresponding control has the form
\begin{eqnarray}
u(t)&=&\frac{i}{T} \tr \big( e^{i H_{0}t}\, V\,e^{-i H_{0}t} L(\frac{t}{T}) \big) + O(\frac{1}{T^{2}})  \nonumber  \\
&=&\frac{i}{T} \sum_{kl}V_{kl}\, e^{i\omega_{kl} t}
L_{lk}(\frac{t}{T}) + O(\frac{1}{T^{2}}). \label{control approx 2}
\end{eqnarray}
The set of pairs $(\psi_0,\{ p_i \})$ of initial states and
target populations for which a solution of \tp (II) provides a
solution of (I) according to (\ref{state approx 2}) and
(\ref{costate approx 2}) is open and full measure in the
corresponding product manifold.

\vspace*{.5 em} According to this theorem, solutions to the \tp
(II) approximate solutions to (I), for large transfer times, in
the sense of equations (\ref{state approx 2}) - (\ref{control
approx 2}), for almost every population transfer. In other words,
all the local minima of \otp (II) approximate local minima for the
\otp (I) according to (\ref{state approx 2}) - (\ref{control
approx 2}). The question arises naturally: Are \emph{all} local
minima of \otp (I), for large $T$, approximated in the sense of
equations (\ref{state approx 2}) - (\ref{control approx 2}) by
local minima of (II)? The answer to this question is essentially
yes (see comment after theorem 3). We state the following theorem:

\vspace*{.5 em} \noindent {\bf{Theorem 3:}} Let $\psi_0$ and $\{
p_i \}_{i=1,\ldots,N}$ be an initial state and a target population
of system (\ref{simple controlled Schrodinger}), respectively.
Then, for almost all pairs $(\psi_0,\{ p_i \})$ and for large
enough $T$, the globally optimal solution to the \otp (I)  is
approximated by the globally optimal solution of (II) according to
theorem 2. As before, the set of pairs $(\psi_0, \{ p_i \})$ of
initial states and target populations for which this happens is
open and full measure in the corresponding product manifold.

\vspace*{.5 em} In fact, we prove that, for large enough $T$, a
number of the lowest energy optima of (I) that depends (in an
unknown way) on $T$, are approximated by the corresponding lowest
energy optima of (II) according to theorem 2. We think though,
that theorem 3 is enough to demonstrate the spirit of our
approach.

Theorem 3 precisely states the main results of our work that were
delineated in the introduction: We obtain useful, physically
plausible properties of the optimal control and state trajectory
and at the same time, we reduce the solution of the original
optimal control problem to a much easier problem: Indeed, the
evolution equations (\ref{simple averaged nl oc state}) and
(\ref{simple averaged nl oc costate}) of \tp (II) do not contain
the free dynamics of the system and thus their solution (which
describes the ``mean'' evolution of the state) is much easier, see
section \ref{Examples}.

\noindent We end this section with two remarks:
\begin{enumerate}
    \item Theorems 2 and 3 are proven for an open, full measure
set of pairs $(\psi_0, \{ p_i \})$ of initial states and target
populations. Unfortunately, the very important case of the initial
state $\psi_0$ being an eigenstate (or, in general, having some
populations equal to $0$) is excluded. The reason is that in this
case, \tp (II) has non-isolated solutions. Although this property
is necessary for our proof of these results, we believe that they
can be extended to (at least some) transfers with non-isolated
solutions. Nevertheless, theorem (I) which contains all the
essential \emph{applicable} aspects of this work still holds.
    \item An interesting implication of the theorem above is
that, for large $T$, the locally optimal values of the objective
(``energy'') scale like $\frac{1}{T}$. This demonstrates that the
quadratic objective we use, cannot correspond to a physical notion
of energy. It could be that an objective like
\[ \int_{0}^{T}  |u(t)| \,dt,\]
would be more appropriate for that purpose in the context of
(\ref{simple controlled Schrodinger}).
\end{enumerate}

\section{Proof of Main Results}
\label{Proof of Main Results}

\subsection*{Proof of Theorem 1}

To begin, we define a new costate variable by $\tilde{\lambda} := T\lambda$ and rewrite (\ref{oc nl state}) and (\ref{oc nl costate}) in terms of $\tilde{\lambda}$:
\begin{eqnarray}
i \dot{\psi} &=& H_{0}\psi +  \frac{i}{T}\,(\tilde{\lambda}^{*} V\psi - \psi^{*}V \tilde{\lambda})\, V\psi,
\label{scaled oc nl state}  \\
i \dot{\tilde{\lambda}} &=& H_{0}\tilde{\lambda} +  \frac{i}{T}\, (\tilde{\lambda}^{*} V\psi - \psi^{*}V \tilde{\lambda})\, V \tilde{\lambda}.
\label{scaled oc nl costate}
\end{eqnarray}
Note that the form of the boundary conditions remains unchanged,
as well:
\begin{equation} \label{scaled oc boundary conditions}
\psi(0)=\psi_{0}, \ \ \  |\psi_{i}(T)|^{2}=p_{i}, \ \ \
\im(\psi^{*}_{i}(T)\,\tilde{\lambda}_{i}(T))=0.
\end{equation}
We must show that $ e^{-i H_{0}t}\,\bar{x}(\frac{t}{T})$ and
$e^{-iH_{0}t} \,\bar{z}(\frac{t}{T})$ satisfy (\ref{scaled oc nl
state}) - (\ref{scaled oc boundary conditions}) up to terms of
order $O(\frac{1}{T})$. To this purpose, we perform one more
change of variables in (\ref{scaled oc nl state}) - (\ref{scaled
oc boundary conditions}):
\[ y = e^{i H_{0}t}\psi, \  \  \  \  l = e^{i H_{0}t} \tilde{\lambda}.  \]
In terms of the new state and costate, the necessary conditions of
optimality take the form
\begin{eqnarray}
i \dot{y} &=& \frac{i}{T}\,(l^{*}F(t)y-y^{*}F(t)l)\, F(t)y,
\label{scaled oc nl state ip} \\
i \dot{l} &=& \frac{i}{T}\,(l^{*}F(t)y-y^{*}F(t)l)\, F(t)l,
\label{scaled oc nl costate ip}
\end{eqnarray}
where, as before, $F(t)=e^{i H_{0}t}\, V\,e^{-i H_{0}t}$, along
with
\begin{equation} \label{scaled oc bcs ip}
y(0)=\psi_{0}, \ \ \  |y_{i}(T)|^{2}=p_{i}, \ \ \
\im(y^{*}_{i}(T)\,l_{i}(T))=0.
\end{equation}
The boundary conditions retain their form because $|y_{i}(T)|^{2}=
|\psi_{i}(T)|^{2}$ and, $l_i(T)$ and $y_i(T)$ are rotated by the
same amount, $e^{-i E_i T}$, with respect to $\psi_i(T)$ and
$\tilde{\lambda}_i(T)$. We will refer to equations (\ref{scaled oc
nl state ip}) - (\ref{scaled oc bcs ip}) as \tp (I'). The
equivalence of problems (I) and (I') is evident.

Now, we have to show that $\bar{x}(\frac{t}{T})$ and
$\bar{z}(\frac{t}{T})$ satisfy (\ref{scaled oc nl state ip}) -
(\ref{scaled oc bcs ip}) up to terms of order $O(\frac{1}{T})$. To
do this, we average equations (\ref{scaled oc nl state ip}) and
(\ref{scaled oc nl costate ip}). To make the procedure more
transparent, we rewrite equations (\ref{scaled oc nl state ip})
and (\ref{scaled oc nl costate ip}) in component form:
\begin{eqnarray*}
\dot{y}_{i}&=&\frac{1}{T} \Big(\sum_{km}V_{km}\, e^{i
\omega_{km} t}
(l_{k}^{*}y_{m}-y_{k}^{*}l_{m}) \Big)
\sum_{j} V_{ij}\, e^{i \omega_{ij} t} y_{j}  \\
\dot{l}_{i}&=&\frac{1}{T} \Big(\sum_{km}V_{km}\, e^{i
\omega_{km} t}
(l_{k}^{*}y_{m}-y_{k}^{*}l_{m}) \Big)
\sum_{j} V_{ij}\, e^{i \omega_{ij} t} l_{j}
\end{eqnarray*}
One can see (based on our controllability assumption) that we get
non-zero contributions from two groups of terms: Terms with
$\omega_{km} \neq 0$ and $\omega_{ij} \neq 0$ such that $m=i$ and
$k=j$, and terms with $\omega_{km}=\omega_{ij}=0$, i.e. $k=m$ and
$i=j$. Letting $\bar{y}$ and $\bar{l}$ be the averaged $y$ and
$l$, the averaged state and co-state equations are:
\begin{eqnarray}
\dot{\bar{y}}_{i}&=& \frac{1}{T} \bigg\{ \sum_{j \neq i}|V_{ij}|^{2}
(\bar{y}_{i}\bar{l}_{j}^{*}- \bar{l}_{i}\bar{y}_{j}^{*})\, \bar{y}_{j}
+ \sum_{k}V_{kk} (\bar{y}_{k}\bar{l}_{k}^{*}- \bar{l}_{k}\bar{y}_{k}^{*})\, V_{ii} \bar{y}_{i} \bigg\},
\label{scaled oc nl state ip in components} \\
\dot{\bar{l}}_{i}&=& \frac{1}{T} \bigg\{ \sum_{j \neq i}|V_{ij}|^{2} (\bar{y}_{i}\bar{l}_{j}^{*}- \bar{l}_{i}\bar{y}_{j}^{*})\, \bar{l}_{j}
+ \sum_{k}V_{kk} (\bar{y}_{k}\bar{l}_{k}^{*}- \bar{l}_{k}\bar{y}_{k}^{*})\, V_{ii} \bar{l}_{i} \bigg\}.
\label{scaled oc nl costate ip in components}
\end{eqnarray}
To finish the proof, we rescale time in equations (\ref{scaled oc
nl state ip in components}) and (\ref{scaled oc nl costate ip in
components}) to $s=\varepsilon t =\frac{t}{T}$. Letting
$\tilde{y}(s) := \bar{y}(t)$ and $\tilde{l}(s) :=
\bar{l}(t)$, these equations read now:
\begin{eqnarray}
\frac{d\tilde{y}_{i}}{ds}&=&   \sum_{j \neq i}|V_{ij}|^{2}
(\tilde{y}_{i}\tilde{l}_{j}^{*}- \tilde{l}_{i}\tilde{y}_{j}^{*})\, \tilde{y}_{j}
+ \sum_{k}V_{kk} (\tilde{y}_{k}\tilde{l}_{k}^{*}- \tilde{l}_{k}\tilde{y}_{k}^{*})\, V_{ii} \tilde{y}_{i},
\label{time scaled oc nl state ip in components} \\
\frac{d\tilde{l}_{i}}{ds}&=&   \sum_{j \neq i}|V_{ij}|^{2} (\tilde{y}_{i}\tilde{l}_{j}^{*}- \tilde{l}_{i}\tilde{y}_{j}^{*})\, \tilde{l}_{j}
+ \sum_{k}V_{kk} (\tilde{y}_{k}\tilde{l}_{k}^{*}- \tilde{l}_{k}\tilde{y}_{k}^{*})\, V_{ii} \tilde{l}_{i}.
\label{time scaled oc nl costate ip in components}
\end{eqnarray}
These equations are the same as (\ref{averaged nl oc state}) and
(\ref{averaged nl oc costate}) (with the substitution $\tilde{y}
\rightarrow \bar{x}$ and  $\tilde{l} \rightarrow \bar{z}$). Then,
from the sequence of variable changes and the averaging theorem,
the conclusion of the theorem follows.$\blacksquare$

\subsection*{Proof of Theorem 2}

To prove Theorem 2 we need the following lemma:
\vspace*{.5 em}

\noindent {\bf{Lemma:}} Let $\psi_0$ and $\{ p_i
\}_{i=1,\ldots,N}$ be an initial state and a target population of
system (\ref{simple controlled Schrodinger}), respectively, and
let $(\bar{x}(s),\, \bar{z}(s))$ be a solution of \tp (II) over
$[0,1]$. The set of pairs $(\psi_0,\{ p_i \})$ of initial states
and target populations for which all solutions of \tp (II) are
\emph{isolated} is open and full measure in the corresponding
product manifold.
\vspace*{.5 em}

\noindent {\bf{Proof of Lemma:}} We begin by introducing new
coordinates for the state and costate of the optimal transfer
problem (II) by
\begin{equation} \label{action-angle}
\bar{x}_i := I_i \, e^{i\phi_i}, \ \ \ \ \bar{z}_i := J_i \, e^{i\theta_i},
\end{equation}
where $\phi_i := \arg \bar{x}_i  \mod \pi$ and $I_i := |\bar{x}_i|$ for $\im\bar{x}_i \geq 0$ and $I_i := -|\bar{x}_i|$ for $\im\bar{x}_i < 0$, and similarly for the $\theta_i$'s and $J_i$'s. Then,
 $I_i, \, J_i  \in \mathbb{R}$ and $\phi_i, \, \theta_i \in [0,\pi),\,
\forall i=1,\ldots,N$. With this definition, the phases $\phi_i$
and $\theta_i$ have discontinuities whenever the signs of the
imaginary parts of $\bar{x}_i$ and $\bar{z}_i$ change. We shall
see that this will not be a problem for us because $\phi_i$ and
$\theta_i$ will turn out to be constant in time. On the other
hand, the introduction of these coordinates will prove to be
beneficial in the following.

In the new coordinates, (\ref{simple averaged nl oc state}) and
(\ref{simple averaged nl oc costate}), take the form
\begin{eqnarray}
i \dot{I}_i - I_i \dot{\phi}_i &=&+ i I_i \, \sum_{j \neq i} |V_{ij}|^2 I_j J_j
\, e^{i(\phi_j - \theta_j)} -i J_i \, e^{i(\theta_i - \phi_i)}
\, \sum_{j \neq i} |V_{ij}|^2 I_j^2, \ \ \ \ \ \
\label{simple averaged nl oc state in action-angle} \\
i \dot{J}_i - J_i \dot{\theta}_i &=& -i J_i \, \sum_{j \neq i} |V_{ij}|^2 I_j J_j \, e^{i(\theta_j - \phi_j)} +i I_i \, e^{i(\phi_i - \theta_i)}
\, \sum_{j \neq i} |V_{ij}|^2 J_j^2, \ \ \ \ \ \
\label{simple averaged nl oc costate in action-angle}
\end{eqnarray}
while the boundary conditions at the end become
\begin{equation}
I_i^2(1)=p_i,  \ \ \ \ \ \ \ \theta_i(1) - \phi_i(1) = 0.
\label{averaged oc boundary conditions in action-angle}
\end{equation}
Multiply (\ref{simple averaged nl oc state in action-angle}) by
$J_i$, (\ref{simple averaged nl oc costate in action-angle}) by
$-I_i$, add them  and take the real part. The resulting equation
reads:
\begin{equation} \label{state-costate relative phases evolution}
I_i J_i (\phi_i - \theta_i)^{.} = \sin(\phi_i - \theta_i) \,
\{I_i^2 \, \sum_{j \neq i} |V_{ij}|^2 J_j^2 -
J_i^2 \, \sum_{j \neq i} |V_{ij}|^2 I_j^2 \}.
\end{equation}
Except for the degenerate cases $(I_i(s)=0,\, J_i(s)=const.)$ and $(J_i(s)=0,\, I_i(s)=const.) \ \forall s \in
[0,1]$, which will be excluded, we see that, given the
transversality conditions at $s=1$, this equation implies that
$\theta_i(s) = \phi_i(s),  \ \forall s \in [0,1]$. Using this
fact, we see that the right sides of equations (\ref{simple
averaged nl oc state in action-angle}) and (\ref{simple averaged
nl oc costate in action-angle}) are purely imaginary. This leads
to the simplified equations
\begin{eqnarray*}
\dot{I}_i  &=&+  I_i \, \sum_{j \neq i} |V_{ij}|^2 I_j J_j
- J_i \, \sum_{j \neq i} |V_{ij}|^2 I_j^2, \\
\dot{J}_i &=& - J_i \, \sum_{j \neq i} |V_{ij}|^2 I_j J_j
+ I_i \, \sum_{j \neq i} |V_{ij}|^2 J_j^2, \\
\dot{\phi}_i(s)&=&0, \ \ \ \ \ \dot{\theta}_i(s)=0.
\end{eqnarray*}
We see that the \tp (II) separates nicely into two parts: A
trivial problem for the arguments of state and costate components
\begin{eqnarray}
\dot{\phi}_i(s)&=&0, \ \ \ \ \ \dot{\theta}_i(s)=0,
\label{simple angle evolution} \\
\phi_i(0)&=&\phi_{i0}, \ \ \theta_i(1) - \phi_i(1) = 0,
\label{averaged oc boundary conditions for angles}
\end{eqnarray}
which has a unique solution as long as $\psi_{0i} \neq 0,\ \forall
i=1, \ldots, N$ (so that all $\phi_i(0)$ are unambiguously
defined) and a \tp for the (signed) magnitudes of state and
costate components
\begin{eqnarray}
\dot{I}_i  &=&+  I_i \, \sum_{j \neq i} |V_{ij}|^2 I_j J_j
- J_i \, \sum_{j \neq i} |V_{ij}|^2 I_j^2,
\label{simple averaged nl oc state in action} \\
\dot{J}_i &=& - J_i \, \sum_{j \neq i} |V_{ij}|^2 I_j J_j
+ I_i \, \sum_{j \neq i} |V_{ij}|^2 J_j^2,
\label{simple averaged nl oc costate in action} \\
I_i(0)&=&I_{i0}, \ \ \ \ I_i^2(1)=p_i.
\label{averaged oc boundary conditions for actions}
\end{eqnarray}

We will refer to (\ref{simple averaged nl oc state in action}) -
(\ref{averaged oc boundary conditions for actions}) as the real
form of problem (II). These \tps can be seen to be necessary
conditions for optimal population transfers of the following
\emph{real} control system on $S^{N-1}$:
\vspace*{.5 em}
\begin{equation} \label{real averaged system}
\frac{dI}{ds}= \left(
\begin{array}{ccc}
0 & |V_{12}|v_{12} &  \cdots \\
-|V_{21}|v_{21}  & 0 & \cdots  \\
\vdots  & \vdots & \ddots
\end{array}
\right) \, I.
\end{equation}
Here the $v_{ij}$ are real controls and (\ref{real averaged
system}) is controllable on $S^{N-1}$ because of the
controllability assumption on the original system, see Appendix B.
This separation of the \tp (II) into a trivial problem for the
evolution of the arguments of state and costate components and
real \tp (II) was inspired by the result of \cite{boscha03}, see
section $4$ of that reference.

We are now ready to prove the lemma. From the discussion so far,
it should be obvious that we need $\psi_{i0} \neq 0$ and
$|\psi_{0i}|^2 \neq p_{i}, \forall i=1, \ldots, N$ so that the
(constant) phases of the state components are uniquely defined and
certain pathological cases mentioned are excluded. The set of
pairs $(\psi_0,\{ p_i \})$ of initial states and target
populations for which this is the case is open and full measure in
the corresponding product manifold.

We define a terminal condition function $\mathcal{G}$ of the
initial costate vector $J(0)$ of real problem (II) by
\begin{equation} \label{real terminal condition function}
\mathcal{G}(J(0)) := (I_2(1)^2, \ldots, I_N(1)^2)^{T}.
\end{equation}
$\mathcal{G}$ is really a function of only $N-1$ components of
$J(0)$ because one of them is fixed by the condition
$0=\bar{x}^{*}(0)\bar{z}(0)=I^{T}(0)J(0)$ (let us decide to fix
$J_1(0))$. Similarly, once $(I_2(1)^2, \ldots, I_N(1)^2)$ are
fixed, so is $I_1(1)^2$, and that is why we need only $N-1$
terminal conditions. In the following, when talking about initial
costate vectors, we will identify vectors in $\mathbb{R}^{N-1}$
with vectors in $\mathbb{R}^{N}$ perpendicular to $I_0$.
$\mathcal{G}$ has two important properties: First, it is a smooth
function of its argument. Indeed, the right sides of (\ref{simple
averaged nl oc state in action}) and (\ref{simple averaged nl oc
costate in action}) are $C^2$ functions of $I$ and $J$ so,
\cite{arn73} their solutions depend smoothly on $I(0)$ and $J(0)$.
Thus, when $I(0)$ is fixed, the terminal conditions depend
smoothly on $J(0)$. Second, because of controllability of system
(\ref{real averaged system}), $\mathcal{G}$ is onto. Since
$\mathcal{G}$ is a mapping between manifolds of the same
dimension, it is a diffeomorphism from (open) neighborhoods of
$\mathbb{R}^{N-1}$ to neighborhoods of the $N-1$ dimensional
simplex. Sard's theorem implies that the set of critical
values of $\mathcal{G}$ in the $N-1$ dimensional simplex (i.e. the
set of points around which $\mathcal{G}$ is not a diffeomorphism)
is of measure zero.

Any initial costate vector $J(0)$ such that $\mathcal{G}(J(0))=
(p_1,\ldots,p_N)^{T}$ provides a solution to real problem (II).
Because of the two properties of $\mathcal{G}$, the set of $p_i$'s
for which \emph{all} $J(0)$ that satisfy $\mathcal{G}(J(0))
=(p_1,\ldots,p_N)^{T}$ satisfy also
$\mathcal{N}(D\mathcal{G}(J(0))) = \emptyset$, is an open set of
full measure in $S^{N-1}$. For a pair of initial state and final
populations which satisfy this, the set of $J(0)$'s must be
discrete and without limit points, otherwise,
$\mathcal{N}(D\mathcal{G}(J(0))) \neq \emptyset$ for at least one
of the $J(0)$'s. This means that, given an initial state with all
populations non-zero, the set of populations that can be achieved
by \emph{isolated} locally optimal transfers of the real problem
(II) is open and full measure. So, the set of pairs
$(\psi_0,\{p_i\})$ of initial states and target populations for
which all solutions of \tp (II) are \emph{isolated} is open and
full measure in the corresponding product manifold.$\blacksquare$
\vspace*{.5 em}

\noindent {\bf{Proof of Theorem 2:}} We are going to define
terminal condition functions for \tps (II) and (I') in a way
similar to that in the proof of the lemma. First, for (II) we
define $\mathcal{F}: \mathbb{R}^{2N-2} \longrightarrow
\mathbb{R}^{2N-2} $ by
\begin{equation}   \label{terminal condition function}
\mathcal{F}(\bar{z}(0)) := (|\bar{x}_2^{2}(1)|,\ldots,|\bar{x}_N^{2}(1)|,\im(\bar{x}^{*}_{2}(1)\,
\bar{z}_{2}(1)),\ldots,\im(\bar{x}^{*}_{N}(1)\,\bar{z}_{N}(1)))^{T}
\end{equation}
(Recall that one complex component of $\bar{z}(0)$ is fixed by
$\bar{x}^{*}(0)\bar{z}(0)=0$. Here, we again identify initial
costate vectors in $\mathbb{C}^{N}$ perpendicular to $\psi_0$ with
vectors in $\mathbb{C}^{N-1}$ and also identify $\mathbb{C}^{N-1}$
with $\mathbb{R}^{2N-2}$). $\mathcal{F}$ is a smooth function of
its argument. Also,  any initial costate vector $\bar{z}(0)$ such
that $\mathcal{F}(\bar{z}(0))=(p_2,\ldots,p_N,0,\ldots,0)^{T}$
provides a solution to problem (II).

The corresponding terminal condition function $\mathcal{F}_1$ for
problem (I'), is defined exactly the same way:
\begin{equation}
\mathcal{F}_1(l(0);T) := (|y_2(T)|^2 ,\ldots,|y_N(T)|^2 ,\im(y^{*}_{2}(T)\,l_{2}(T)),\ldots,\im(y^{*}_{N}(T)\,l_{N}(T)))^{T}
\label{terminal condition function for problem I'}
\end{equation}
The second argument of $\mathcal{F}_I$ is just a reminder of the
transfer time. $\mathcal{F}_1$ is also a smooth function of its
argument. Again, any initial costate vector $y(0)$ such that
$\mathcal{F}_1(y(0))=(p_2,\ldots,p_N,0,\ldots,0)^{T}$ provides a
solution to problem (I'). From the proof of theorem (I), we know
that
\begin{equation} \label{relation of F's for problems I' and II}
\mathcal{F}_1(v;T)= \mathcal{F}(v)   + O(\frac{1}{T}).
\end{equation}
From this we also have that
\[ D\mathcal{F}_1(v;T)= D\mathcal{F}(v) + O(\frac{1}{T}). \]
Although $\mathcal{F}_1$ is not formally defined for $T=\infty$
from (\ref{terminal condition function for problem I'}), we can
define it from (\ref{relation of F's for problems I' and II}) as
$\mathcal{F}_1(v;0) := \mathcal{F}(v)$. With this definition,
$\mathcal{F}_1$ is continuous in $\frac{1}{T}$ at $\frac{1}{T}=0$,
with continuous first derivatives in $v$ and $\frac{1}{T}$ there.
In particular, $ D\mathcal{F}_1(v;\infty)= D\mathcal{F}(v)$.

Consider now an initial costate vector $v$ that solves problem
(II), i.e.
\[ \mathcal{F}(v) = (p_2,\ldots,p_N,0,\ldots,0)^{T}\]
We have seen that  the set of of pairs $(\psi_0,\{ p_i \})$ such
that \emph{all }$v$'s that satisfy this are isolated
($D\mathcal{F}(v)$ full rank), is an open set of full measure in
the product space. For such a transfer and for large enough $T$,
the implicit function theorem guarantees the existence of a
$\delta v$ such that $\mathcal{F}_1(v+\delta v;T)=
(p_2,\ldots,p_N,0,\ldots,0)^{T}$. Then, $v+\delta v$ provides a
solution for problem (I') and $\frac{v+\delta v}{T}$ is a solution
for problem (I). Taking $T$ large enough so that $|\delta
v|=O(\frac{1}{T})$, the averaging theorem guarantees that this
solution $(\psi(t),\lambda(t))$ to \tp (I) is such that
\begin{eqnarray*}
\psi(t)&=&e^{-i H_{0}t}\,\bar{x}(\frac{t}{T}) + O(\frac{1}{T}), \\
\lambda(t)&=& \frac{1}{T} e^{-i H_{0}t}\,\bar{z}(\frac{t}{T})+ O(\frac{1}{T^{2}}),
\end{eqnarray*}
where $(\bar{x}(s),\bar{z}(s))$ is the solution to problem (II) we
started with.$\blacksquare$

\subsection*{Proof of Theorem 3}

We only consider transfers such that problem (II) has isolated solutions. Any initial costate $v$ that satisfies $\mathcal{F}(v)=(p_2,\ldots,p_N,0,\ldots,0)^{T}$
provides a solution to problem (II) and $D\mathcal{F}(v)$ is full
rank. Inside a ball of radius $M>0$ there can be only a
\emph{finite} number of these initial costates $v$ because any
discrete set with no limit points inside a compact set must be finite.

From equation (\ref{relation of F's for problems I' and II}) we
may conclude that for a given transfer, we can take $T$ large
enough to bound the difference of $\mathcal{F}$ and
$\mathcal{F}_1$ over an open ball around the origin by any 
$\delta>0$:
\[ \forall \, M>0,\ \forall \, \delta>0,\  \exists \, T>0 \ \mathrm{such \ that} 
\ |\mathcal{F}_1(v;T)-\mathcal{F}(v)|<\delta, \ \forall \, |v|<M.  \]
Thus, the only solutions of $\mathcal{F}_1(\tilde{v};T)=(p_2,
\ldots, p_N, 0, \ldots,0)^{T}$ inside the ball of radius $M$ come
from perturbing solutions of $\mathcal{F}(v)=(p_2, \ldots, p_N, 0,
\ldots, 0)^{T}$ by quantities of order $O(\frac{1}{T})$, for $T$
large enough. In particular, they also form a finite set and
$D\mathcal{F}_1(\tilde{v};T)$ is full rank for each such
$\tilde{v}$. We arrive at exactly the same conclusion if, instead
of a ball, we define a neighborhood of the origin by an ellipsoid.

We introduce now the following quadratic form in $v \in \{w \in
\mathbb{C}^{N}\ \mathrm{s.t.} \ \psi_0^{*}w=0\} \simeq
\mathbb{C}^{N-1}\simeq \mathbb{R}^{2N-2}$:
\[ E(v)= \sum_{i \neq j}^{N} |V_{ij}|^2 |\psi_{0i} v_j^{*}-v_i \psi_{0j}^{*} |^2. \]
$E$ is non-negative and, furthermore, due to the connectivity of
the graph of $V$ (part of the controllability assumption) can be
shown to be positive definite, see Appendix C. Then, the sub-level
sets of $E$ define (open) ellipsoids in $\mathbb{R}^{2N-2}$. The
significance of our choice for $E(v)$ lies in the following:
\[ H(\bar{x},\bar{z})=
\sum_{i \neq j}  |V_{ji}|^2 |\bar{x}_{i}\bar{z}_{j}^{*}- \bar{z}_{i}\bar{x}_{j}^{*}|^2  \]
is the Hamiltonian function from which the optimal state and
costate equations (\ref{simple averaged nl oc state}) and
(\ref{simple averaged nl oc costate}) are derived. Thus, $H$ is a
constant of motion along the optimal solutions. Note, also, that
$H(\bar{x},\bar{z})=\sum_{i \neq j} |u_{ij}|^{2}$. Since $E$ is
just $H$ evaluated at $t=0$, we conclude that $E$ is equal to the
cost of a trajectory of system (II) (equations (\ref{simple
averaged nl oc state}) and (\ref{simple averaged nl oc
costate})\,) with initial conditions $(\psi_0,v)$:
\[ \int_{0}^{1}\sum_{i \neq j} |u_{ij}(s)|^{2} ds=H(0)=\sum_{i \neq j}^{N} |V_{ij}|^2 |\psi_{0i} v_j^{*}-v_i \psi_{0j}^{*} |^2. \]
With a calculation similar to that of section \ref{Optimal
population transfers for an averaged system}, one can show that
the locally optimal costs for problems (II) and (I/I') coming from
the solutions $v$ and $\tilde{v}=v+O(\frac{1}{T})$, respectively
are related as follows:
\begin{eqnarray*}
\int_{0}^{T}  u^{2}(t) \,dt & = &  \frac{1}{T} \big( \int_{0}^{1}
\sum_{i \neq j} |u_{ij}(s)|^2  \, ds + O(\frac{1}{T})\, \big) \\
 & = &\frac{1}{T} \big( \sum_{i \neq j}^{N} |V_{ij}|^2 |\psi_{0i} v_j^{*}-v_i \psi_{0j}^{*} |^2 + O(\frac{1}{T})\, \big).
\end{eqnarray*}
Let us fix a value $E_0>0$ such that the initial costate $v_0$ that
achieves the desired transfer with the \emph{minimum} energy for
problem (II) satisfies
\[ \sum_{i \neq j}^{N} |V_{ij}|^2 |\psi_{0i} v_j^{*}
-v_i \psi_{0j}^{*} |^2 < E_{0}. \]
Then, for large enough $T$, $\tilde{v}_0=v_0 + O(\frac{1}{T})$ is
the initial costate that achieves the desired transfer with the
\emph{minimum} energy for  problem (I'). This proves the assertion
of the theorem. In fact, we proved a little bit more: Not only the
global optimum, but also all local optima of problem (I') with
values of energy  less than $\frac{E_0}{T}$, come from local
optima of problem (II) according to the theorem 2, for $T$ large
enough (Note that the solutions $\tilde{v}$ of problem (I/I')
\emph{outside} the ellipsoid $E(\tilde{v})<E_0$ have higher costs
than those inside the ellipsoid).$\blacksquare$

\section{Examples}
\label{Examples}

We consider three examples, each involving a different quantum
system: A general two-state system with one control, a general
three-state system with one control and the Morse oscillator model
for the vibrational dynamics of the ground electronic state of the OH bond. We restrict our attention to the 22-dimensional space of bound states for that model and seek to control populations again with one control field. In all of the examples, we consider transfers from one eigenstate of the system, to another. Although in the proof of theorems 2 and 3 we had to exclude such transfers (because in that case \tp (II) has continua of solutions parameterized by angles, see (\ref{simple angle evolution}) and (\ref{averaged oc boundary conditions for
angles}), and we need isolated solutions to prove the theorems),
we believe these theorems to hold for such transfers as well.
Perhaps this can be established using different techniques from
ours. In any case, theorem 1 still holds and so, every solution of
\tp (II) will furnish an approximate solution to (I).

\subsection*{A two-state system}
\label{A two-state system}

Consider the two-level system $i\dot{\psi}=(H_0 + Vu)\psi$, $\psi
\in \mathbb{C}^{2}$, with
\[ H_{0}= \left( \begin{array}{cc} E_1&0 \\ 0&E_2 \end{array} \right),
\ \ \mathrm{and} \ \
V= \left(\begin{array}{cc} V_{11}&V_{12} \\ V_{12}^{*}&V_{22}
\end{array} \right). \]
By rescaling $u$, we  make $|V_{12}|=1$. We are
interested in the ``population inversion'' transfer
\[ \psi_0 = \left(%
\begin{array}{c}
  1 \\
  0 \\
\end{array}%
\right) \ \ \ \longrightarrow \ \ \ \psi_d = \left(%
\begin{array}{c}
  0 \\
  1 \\
\end{array}%
\right).\]
In this example, the averaged \otp (II) can be solved analytically. To begin, we introduce the anti-Hermitian matrices $L$ and $K(L)$ of
section (\ref{Optimal population transfers for an averaged system}):
\[ L=\left( \begin{array}{rr} 0 & L_{12} \\ - L_{12}^{*} & 0 \\ \end{array}
\right) \ \mathrm{and} \ K(L)=L. \]
Then, (\ref{averaged su(n) costate eq}) implies that $L_{12}(s), \
s \in [0,1]$, is constant. Equation (\ref{averaged oc state2}) reads
\[ \frac{d\bar{x}}{ds}=\left( \begin{array}{rr} 0&L_{12} \\ - L_{12}^{*}&0 \\
\end{array} \right)\bar{x}. \]
Its solution with initial condition $\left( \begin{array}{c}
1\\0\\ \end{array} \right)$, is given by
\[ \bar{x}(s)=\left( \begin{array}{r} \cos(|L_{12}|s) \\ -i \frac{L_{12}}{|L_{12}|}\sin(|L_{12}|s) \end{array} \right). \]
To achieve  $\bar{x}(1)= \left( \begin{array}{c} 0 \\ e^{i*}
\\ \end{array} \right)$, we must have $|L_{12}|=(n+\frac{1}{2})\,\pi$, with $n \in
\mathbb{N}$. The value of the cost (\ref{objective in averaged
system}) is $\frac{\pi^{2}}{2}(2n+1)^{2}$. Thus, $n=0$ corresponds
to the global minimum of \otp (II). The approximate optimal
control for problem (I) has the form (\ref{control approx
2})
\[ u(t)=-\frac{\pi}{T} \sin(\omega_{21}t+\varphi), \]
where $\varphi \in [0,2\pi)$ comes from the phases of $L_{12}$ and
$V_{12}$.

Figure (\ref{oc ave, 2 state,1->2, T=10pi and 6pi,populations}) shows the evolution of populations in a two-state system with $\omega_{21}=1$ under the approximate optimal control, for $T=10\pi$ and $T=6\pi$. Note that even for  $T=6\pi$, the averaged equations are
still a good approximation to the full dynamics.
\begin{figure}[!h]
\begin{center}
\scalebox{.65}{\includegraphics{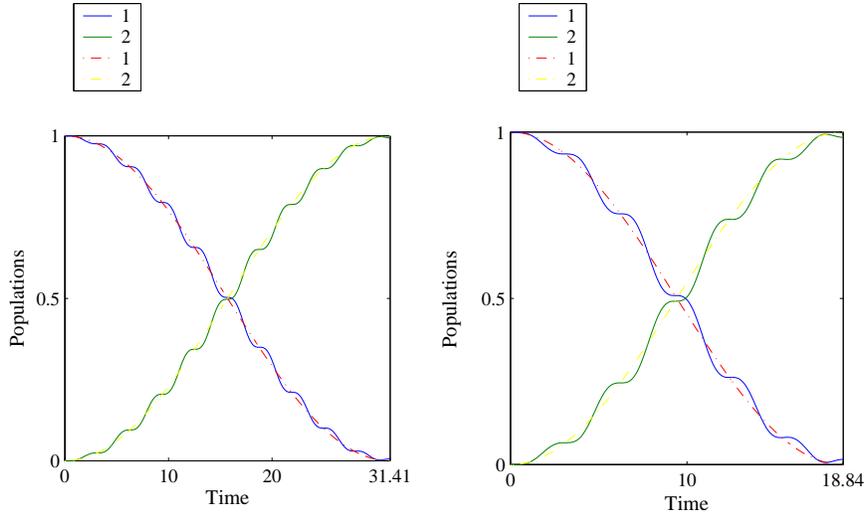}}
\caption{Averaged (dashed line) and exact (full line) populations of
the two-state system under the approximate optimal control, for
$T=10\pi$ and $T=6\pi$.} \label{oc ave, 2 state,1->2, T=10pi and 6pi,populations}
\end{center}
\end{figure}

\subsection*{A three-state system}
Let us now consider the general three-state system with one control and
\[ H_{0}=\left(%
\begin{array}{ccc}
  E_{1} & 0 & 0 \\
  0 & E_{2} & 0 \\
  0 & 0 & E_{3} \\
\end{array}%
\right)
\ \mathrm{and} \
V=\left(%
\begin{array}{ccc}
  V_{11} & V_{12} &V_{13} \\
  V_{12}^{*} & V_{22} & V_{23} \\
 V_{13}^{*} & V_{23}^{*} & V_{33} \\
\end{array}%
\right).  \]
We assume that $\omega_{12}=E_1 - E_2$, $\omega_{13}=
E_1 - E_3$ and $\omega_{23}=E_2 - E_3$ are all different from each other and from zero. Their exact values are unimportant for the averaged \otp as are the values of $V_{11}$, $V_{22}$ and $V_{33}$. By rescaling the control, we can take $|V_{12}|=1$ (we assume $|V_{12}| \neq 0$). Define $p:=|V_{23}|^2$ and $r=|V_{13}|^2$. We assume that $1 >p>r \geq 0$, with other cases treated similarly.  We are interested in the transfer
\[ \psi_{0}=\left(%
\begin{array}{c}
  1 \\
  0 \\
  0 \\
\end{array}%
\right) \ \ \longrightarrow \ \ \psi_{d}=\left(%
\begin{array}{c}
  0 \\
  0 \\
  1 \\
\end{array}%
\right), \]
particularly in the way the ``two-photon'' transition
\[\left(%
\begin{array}{c}
  1 \\
  0 \\
  0 \\
\end{array}%
\right)  \ \longrightarrow \  \left(%
\begin{array}{c}
  0 \\
  1 \\
  0 \\
\end{array}%
\right)  \ \longrightarrow \  \left(%
\begin{array}{c}
  0 \\
  0 \\
  1 \\
\end{array}%
\right) ,\]
assists the ``direct'' (``one-photon'') transition
\[\left(%
\begin{array}{c}
  1 \\
  0 \\
  0 \\
\end{array}%
\right)  \ \longrightarrow  \ \left(%
\begin{array}{c}
  0 \\
  0 \\
  1 \\
\end{array}%
\right). \]
In this example, we are able to calculate the form of the (locally) optimal controls for the averaged problem analytically up to a constant, which has to be computed by solving the state evolution equations numerically and imposing the terminal conditions on the state. For a special case ($p=1$), we can obtain everything analytically. We begin again with equation  (\ref{averaged su(n) costate eq}), $\frac{dL}{ds}=[K(L),L]$. The diagonal elements of $L$ are zero and $L$ is anti-Hermitian, so it has only 3 independent (complex) entries, $L_{12}$, $L_{23}$ and $L_{13}$. They satisfy the following equations:
\begin{eqnarray*}
\frac{dL_{12}}{ds} & = & (p-r) L_{13}L_{23}^{*}, \\
\frac{dL_{23}}{ds} & = & (r-1)L_{12}^{*}L_{13}, \\
\frac{dL_{13}}{ds} & = & (1-p)L_{12}L_{23}.   
\end{eqnarray*}
Because $\bar{x}_2(0)=\bar{x}_3(0)=0$, we have that $L_{32}(0)=0$.
The general solution of the above equations with $L_{32}(0)=0$ is
\begin{eqnarray*}
L_{12}(s)&=&e^{i\phi_{12}}\, A \, \cn (ws), \\
L_{23}(s)&=&-e^{i\phi_{23}}\, B \, \sn (ws), \\
L_{13}(s)&=&e^{i(\phi_{12} + \phi_{23})}\, C \, \dn (ws),
\end{eqnarray*}
where $w>0$. $\cn$, $\sn$ and $\dn$ are Jacobi elliptic functions and
\[ \left(%
\begin{array}{c}
  A \\
  B \\
  C \\
\end{array}%
\right) = \frac{1}{\sqrt{(1-p)(1-r)(p-r)}} \,
\left(%
\begin{array}{c}
  kw\sqrt{p-r} \\
  kw\sqrt{1-r} \\
  w\sqrt{1-p} \\
\end{array}%
\right).\] 
$0 \leq k \leq 1$ is the modulus of the elliptic functions. Now, from $\bar{x}_1(1)=\bar{x}_2(1)=0$, we have that
$L_{12}(1)=0$. This allows us to determine $w$ as $w=(2n+1)K(k)$, $n \in \mathbb{N}$, where 
\[ K(k):=\int_{0}^{\frac{\pi}{2}}\frac{d\theta}{\sqrt{1-k^2 \sin^2\theta}}, \]
the complete elliptic integral of the first kind, is the quarter-period of the functions $\cn$ and $\sn$. The only undetermined parameter is $k$. It can be solved for numerically by  solving the \tp given by (\ref{averaged oc state2}),
\[ \frac{d\bar{x}}{ds}= \left(%
\begin{array}{lcr}
0 & e^{i\phi_{12}}\, A \, \cn (ws) & 
e^{i(\phi_{12}+\phi_{23})}\, rC \, \dn (ws) \\
-e^{-i\phi_{12}}\, A \, \cn (ws) & 0 &
-e^{i\phi_{23}}\, pB \, \sn (ws)  \\
-e^{-i(\phi_{12}+\phi_{23})}\, rC \, \dn (ws) & e^{-i\phi_{23}}\, pB \, \sn (ws)  & 0 \\
\end{array}%
\right) \bar{x}, \]
and the boundary conditions
\[ \bar{x}(0)= \left(%
\begin{array}{c}
  1 \\
  0 \\
  0 \\
\end{array}%
\right), \  \bar{x}(1)= \left(%
\begin{array}{c}
  0 \\
  0 \\
  e^{i*} \\
\end{array}%
\right). \]
It is straightforward to see that the phases $\phi_{12}$, $\phi_{23}$ and $\phi_{12}+\phi_{23}$ can be absorbed in the phases of the components of  $\bar{x}$, so $k$ is independent of them and depends only on $p$ and $r$. Intuitively, one expects a discrete set of solutions for $k$. The cost of a local minimizer is given by 
\begin{eqnarray*}
J&=&2\int_{0}^{1}[|u_{12}(s)|^2+|u_{23}(s)|^2+|u_{13}(s)|^2]\, ds \\
&=&2\int_{0}^{1}(A^2 \cn^2(ws)+rC^2 \dn^2(cs)+pB^2 \dn^{2} (cs))\, ds \\
&=&\frac{2(2n+1)^2 K^2(k)}{(1-p)(1-r)(p-r)}[(p-r)k^2 + r(1-p)].
\end{eqnarray*}
The expression for the approximate locally optimal controls  of \otp (I) is
\begin{eqnarray*}
u(t)=\frac{2}{T}\Big\{-A\, \cn(w\frac{t}{T})&& \hspace*{-1.5em} \sin(\omega_{21}t-\alpha_{12}+\phi_{12})\, + \\
\sqrt{p}\, B\, \sn(w\frac{t}{T})&& \hspace*{-1.5em}  
\sin(\omega_{32}t-\alpha_{23}+\phi_{23})\, - \\
\sqrt{r}\, C\, \dn(w\frac{t}{T})&& \hspace*{-1.5em}  
\sin(\omega_{31}t-\alpha_{13}+\phi_{12}+\phi_{23}) \Big\},
\end{eqnarray*}
where $\alpha_{ij}=\arg V_{ij}$.

Figures (\ref{oc ave,3state,1-->3,T=20pi,control&profiles}) and 
(\ref{oc ave,3state,1-->3,T=20pi,populations}) show an approximate locally optimal control, the (slowly-varying) profiles of its Bohr frequency
components and the averaged and exact evolution of the three state system with $p=.9$ and $r=.1$, under this control for $T=20\pi$. Figure 
(\ref{oc ave,3state,1-->3,T=20pi,populations}), in particular, demonstrates the point we discussed in the introduction: The `averaged' \tp is non-stiff because the short-time scale natural dynamics of the system has been averaged over and thus, its solutions are much easier to compute compared with those of the original \tp (I). In Appendix D, the special case with $p=1$ is analyzed in detail and a complete solution to the \otp (II) is given for that case.
\begin{figure}[!h]
\begin{center}
\scalebox{.6}{\includegraphics{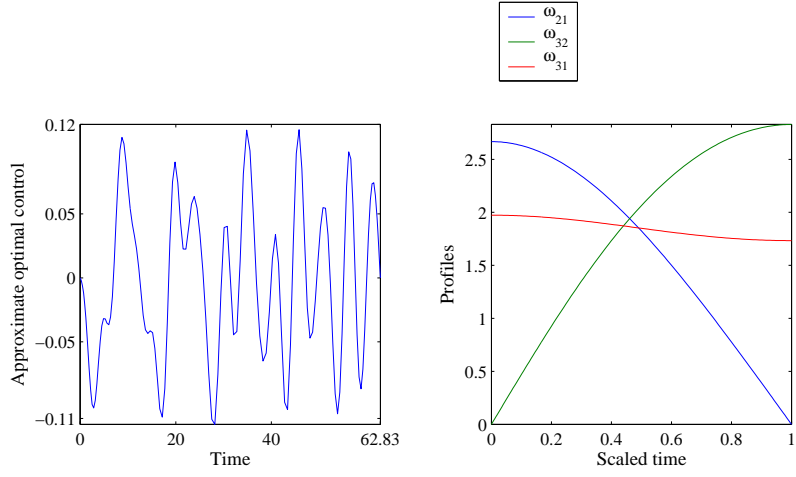}}
\caption{Approximate optimal control for the transition $1
\rightarrow 3$ in $T=20\pi$ and the corresponding frequency profiles.}
\label{oc ave,3state,1-->3,T=20pi,control&profiles}
\end{center}
\end{figure}

\begin{figure}[!h]
\begin{center}
\scalebox{.6}{\includegraphics{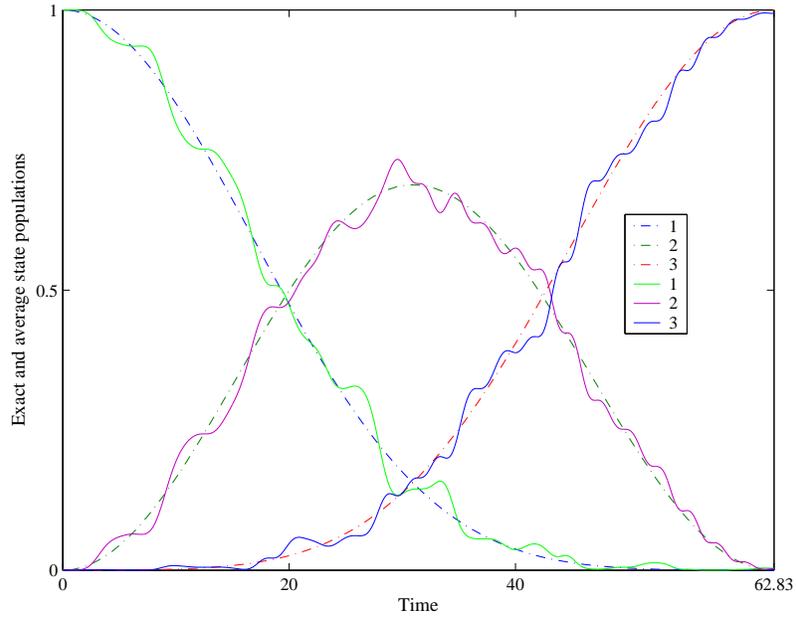}}
\caption{Averaged (dashed lines) and exact (full lines) populations of
the three-state system under the approximate optimal control.}
\label{oc ave,3state,1-->3,T=20pi,populations}
\end{center}
\end{figure}

\subsection*{Bound states of a Morse oscillator}

Our final example considers the $22$-dimensional space of bound states of the Morse oscillator model for the vibrational dynamics of the ground electronic state of the OH molecule. Here we solve the \otp (II) numerically for two transfers, one from the ground vibrational state (state 1) to an intermediate excited vibrational state (state 10) and one from state 10 to the highest bound state, state 22. Figures (\ref{oc ave,Morse,1-->10,average,populations and profiles}) and (\ref{oc ave,Morse,10-->22,average,populations and profiles}) contain the state populations as well as the intensities (absolute values squared) of the frequency profiles as functions of rescaled time. Note the correspondence between high intensity value for a profile $L_{ij}$ and the transition between states $i$ and $j$. 

\begin{figure}[!h]
\begin{center}
\scalebox{.6}{\includegraphics{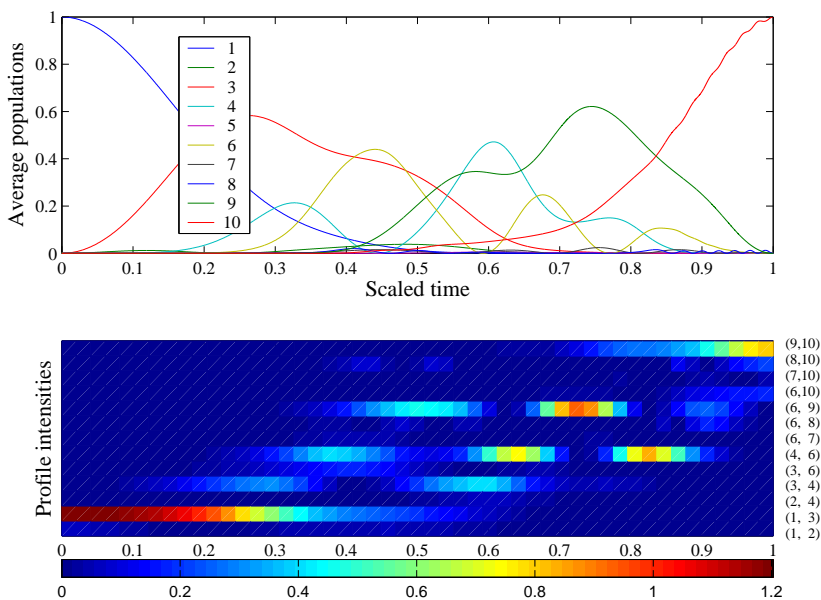}}
\caption{Average populations and profile intensities vs. (scaled) time for a locally optimal transfer $1 \rightarrow 10$.}
\label{oc ave,Morse,1-->10,average,populations and profiles}
\end{center}
\end{figure}

\begin{figure}[!h]
\begin{center}
\scalebox{.6}{\includegraphics{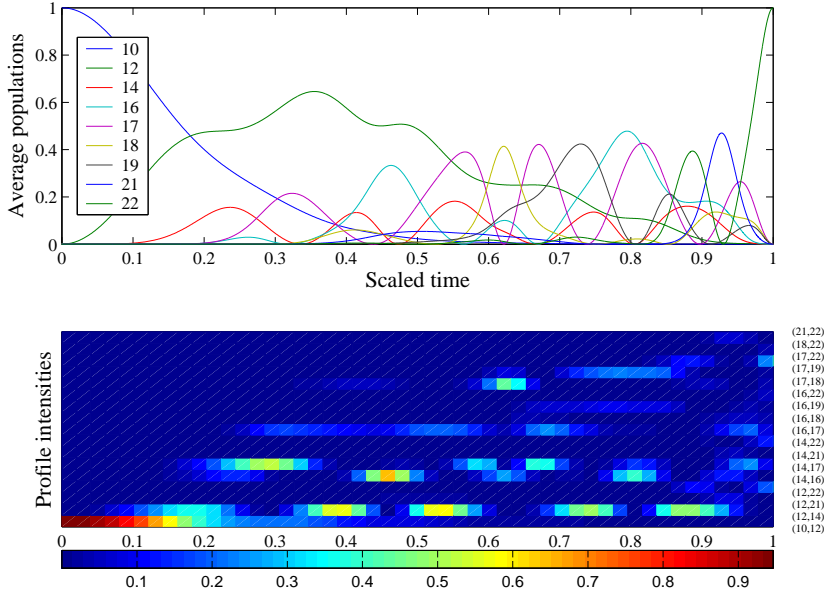}}
\caption{Average populations and profile intensities vs. (scaled) time for a locally optimal transfer $10 \rightarrow 22$.}
\label{oc ave,Morse,10-->22,average,populations and profiles}
\end{center}
\end{figure}

\section{Conclusion}
\label{Conclusion}

We examined the large transfer time limit of exact, optimal
population transfers in a finite dimensional quantum system. The
investigation of this problem uncovered useful structure in the
optimal control and state trajectory and, moreover, resulted in the
much simpler optimal control problem (II) whose solution provides the
first order solution to the original optimal transfer problem, in
a $\frac{1}{T}$ expansion. The main reason we considered \emph{exact} population transfers is that in this case, the `average' \tp (II) (equations (\ref{averaged nl oc state}) - (\ref{averaged oc boundary conditions}) ) is independent of the transfer time $T$ of the original problem. This in turn allowed us to prove Theorems 2 and 3 which are important for uncovering the structure of solutions of \otp (I). Moreover, an advantage of the \otp (I) is that, once one has a solution to the `average'  \tp (II) one can derive from it approximate solutions to \otp (I) \emph{for all} transfer times $T$ larger than some $T_0$. On the other hand, solving the \tp (II) becomes extremely difficult as the dimension of the system grows. For practical applications to large-dimensional systems, an \otp with an objective like (\ref{relaxed cost functional})
\[ a \int_{0}^{T} u^{2}(t) \,dt \, +
\sum_{i=1}^{N}\big(|\psi_{i}(T)|^{2}-p_{i} \big)^2, \]
which leads to \emph{separated} boundary conditions would be preferable. Although an analog of Theorem (I) can be proven in this case too, it doesn't seem so for Theorems 2 and 3. Yet, our technique of averaging over the short scale natural dynamics of the system would very much improve the efficiency of any method used for the solution of these problems, as well.

\section*{Appendix A}

First, we prove the transversality conditions
$\im(\psi^{*}_{i}(T)\,\lambda_{i}(T))=0, \ i=1,\ldots,N$. The \otp
(I) is a standard Bolza problem with (real) terminal state
constraints, $|\psi_{i}(T)|^{2}=p_{i},\ i=1,\ldots,N$. According
to the general theory of such problems \cite{sagwhi77}, the
transversality conditions for the costate at the final time are
given by
\begin{equation}\label{transversality conditions}
\lambda_{i}(T)=\nu_{i} \, \frac{\partial
(|\psi_{i}(T)|^{2}-p_i)}{\partial
\psi_{i}(T)^{*}}=\nu_{i}\psi_{i}(T),
\end{equation}
where $\nu_{i}$ are the real Lagrange multipliers that enforce the
terminal state constraints. The transversality conditions in the
form we state them,
\[ \im(\psi^{*}_{i}(T)\,\lambda_{i}(T)\,)=0, \ i=1,\ldots,N \]
follow easily.

Next, we prove that $\lambda(t)^{*}\psi(t)=0$. One can easily show
from equations (\ref{oc state}) and (\ref{oc costate}) that
$\lambda(t)^{*}\psi(t)$ is constant along any optimal trajectory
and so, $\lambda(t)^{*}\psi(t)=\lambda(T)^{*}\psi(T)$. From the
transversality conditions it follows that
$\im(\psi^{*}(T)\,\lambda(T))=0$ and thus,
$\im(\psi^{*}(t)\,\lambda(t))=0$. Let us now decompose $\lambda$
as follows, $\lambda=c \psi+\lambda_{\perp}$, with $c$ real and
$\lambda_{\perp}$ perpendicular to $\psi$, i.e.
$\lambda_{\perp}^{*}\psi=0$. All we need to show is that $c=0$. We
introduce this representation of $\lambda$ into equation (\ref{oc
control}). The resulting equation is
\[ u = i\,(\lambda_{\perp}^{*}V\psi - \psi^{*}V\lambda_{\perp}) + i\, (c^{*}-c)\, (\psi^{*}V\psi), \]
and the reality of $u$ forces $c$ to be zero.

\section*{Appendix B}

We prove that system (\ref{averaged simple system in [0,1]}) is
controllable on account of the controllability assumption on the
original system (\ref{simple controlled Schrodinger}). Due to the
fact that (\ref{averaged simple system in [0,1]}) has complex
controls, every non-zero $V_{ij}, i \neq j$, provides us with two
generators of $su(N)$ (traceless anti-Hermitian matrices),
$E_{ij}-E_{ji}$ and $i(E_{ij}+E_{ji})$ ($E_{ij}$ denotes a matrix
with only one non-zero element, at the position $(i,j)$, equal to
one. So, $(E_{ij})_{kl}=\delta_{ik}\delta_{jl}$. It is easy to see
that $E_{ij}E_{mn}=\delta_{jm}E_{in}$). One can easily verify the
following commutation relations:
\begin{eqnarray}
[E_{ij}-E_{ji},E_{jk}-E_{kj}] &=& E_{ik}-E_{ki},\ i \neq k,
\label{real generators} \\
E_{ij}-E_{ji},i(E_{jk}+E_{kj}) &=& i(E_{ik}+E_{ki}), \ i \neq k,
\label{real&imaginary generators} \\
E_{ij}-E_{ji},i(E_{ji}+E_{ij}) &=& 2i(E_{ii}-E_{jj}).
\label{real&imaginary generators 2}
\end{eqnarray}
Due to the connectivity of the graph of $V$, there exists a
sequence of index pairs that connects any state index $i$ with any
other state index $j$. Thus, starting with the given generators
$E_{ij}-E_{ji}$ and $i(E_{ij}+E_{ji})$ for all $V_{ij}\neq 0, i
\neq j$, we can generate, with repeated use of (\ref{real
generators}) and (\ref{real&imaginary generators}), all missing
such generators (corresponding to $V_{ij}= 0, i \neq j$). Finally,
using (\ref{real&imaginary generators 2}) we can generate the
diagonal generators of $su(N)$ (its Cartan subalgebra). For the
system (\ref{real averaged system}) which has real controls, every
non-zero $V_{ij}, i \neq j$, provides us with a generator of
$so(N)$ (anti-symmetric matrices), $E_{ij}-E_{ji}$. In that case,
the repeated use of (\ref{real generators}) is enough to establish
controllability.

\section*{Appendix C}

Here we show that the quadratic form
\[ E(v)= \sum_{i \neq j}^{N} |V_{ij}|^2 |\psi_{0i} v_j^{*}-v_i \psi_{0j}^{*} |^2 \]
in $v \in \{w \in \mathbb{C}^{N}\ \mathrm{s.t.} \ \psi_0^{*}w=0\}
\simeq \mathbb{C}^{N-1}\simeq \mathbb{R}^{2N-2}$ is positive
definite, based on the connectivity of the graph of $V$. The
non-negativity of $E$ is obvious. Let $E(v)=0$, then we have that
\begin{equation}\label{E(v)=0}
\psi_{0i} v_j^{*}-v_i \psi_{0j}^{*}=0, \ \forall \ (i,j)\
\mathrm{s.t.} \ V_{ij}\neq 0.
\end{equation}
Let us assume for a moment that $V_{12} \neq 0$. Then (recall, all
$\psi_{0i} \neq 0$),
\[ \frac{v_2}{\psi_{02}}=\frac{v_1^{*}}{\psi_{01}^{*}}. \]
It is straightforward to see that the connectivity of the graph of
$V$ and the repeated use of the relations (\ref{E(v)=0}) allows us
to show that
\[ \frac{v_i}{\psi_{0i}}= \frac{v_1}{\psi_{01}} \ \mathrm{or} \ \frac{v_1^{*}}{\psi_{01}^{*}}, \forall \ i=2,\ldots,N.  \]
Recall now that both $\psi_0$ and $v$ can be defined modulo global
phases which we choose such that $\psi_{01}$ and $v_1$ are real.
Then,
\[ \frac{v_i}{\psi_{0i}}= \frac{v_1}{\psi_{01}}, \forall \ i=2,\ldots,N.  \]

The relation $\psi_0^{*}v=0$ implies then that
\[ (\sum_{i=1}^{N} |\psi_{0i}|^{2})\, \frac{v_1}{\psi_{01}}=0, \]
which means that $v_1=0$ and hence $v=0$.

\section*{Appendix D}

In the special case $p=1$, we can give a complete analytic solution to the \otp (II) for the second example of section \ref{Examples}. The solution for the matrix elements of $L$ is given by
\begin{eqnarray*}
L_{12}(s)&=&e^{i\phi_{12}}\, A \, \cos (ws), \\
L_{23}(s)&=&-e^{i\phi_{23}}\, A \, \sin (ws), \\
L_{13}(s)&=&e^{i(\phi_{12} + \phi_{23})}\, \frac{w}{1-r},
\end{eqnarray*}
where $w$ is determined to be $w=(2n+1)\frac{\pi}{2}$, $n \in \mathbb{N}$. To determine $A$, we have to solve the \tp given by (\ref{averaged oc state2}),
\[ \frac{d\bar{x}}{ds}= \left(%
\begin{array}{lcr}
0 & e^{i\phi_{12}}\, A \, \cos (ws) & 
e^{i(\phi_{12}+\phi_{23})}\, \frac{rw}{1-r} \\
-e^{-i\phi_{12}}\, A \, \cos (ws) & 0 &
-e^{i\phi_{23}}\, A \, \sin (ws)  \\
-e^{-i(\phi_{12}+\phi_{23})}\, \frac{rw}{1-r} &
e^{-i\phi_{23}}\, A \, \sin (ws)  & 0 \\
\end{array}%
\right) \bar{x}, \]
and the boundary conditions
\[ \bar{x}(0)= \left(%
\begin{array}{c}
  1 \\
  0 \\
  0 \\
\end{array}%
\right),  \  \bar{x}(1)= \left(%
\begin{array}{c}
  0 \\
  0 \\
  e^{i*} \\
\end{array}%
\right). \]
With the change of variables
\[ y= \left(%
\begin{array}{c}
  y_1 \\
  y_2 \\
  y_3 \\
\end{array}%
\right) := \left(%
\begin{array}{lcr}
\cos (ws) & 0 & \sin (ws) \\ 
0 & 1 & 0 \\
-\sin (ws) & 0 & \cos (ws) \\
\end{array}%
\right)  \left(%
\begin{array}{r}
 \bar{x}_1 \\
 e^{i\phi_{12}}\, \bar{x}_2 \\
 e^{i(\phi_{12}+\phi_{23})}\,  \bar{x}_3 \\
\end{array}%
\right), \]
$y$ satisfies the simpler equation
\[ \frac{dy}{ds}= \left(%
\begin{array}{lcr}
0 & A & \frac{w}{1-r} \\
-A & 0 & 0 \\
-\frac{w}{1-r} &0 & 0 \\
\end{array}%
\right) y. \]
The boundary conditions for $\bar{x}$ translate into the following conditions for $y$:
\[ y(0)= \left(%
\begin{array}{c}
  1 \\
  0 \\
  0 \\
\end{array}%
\right),  \  y(1)= \left(%
\begin{array}{c}
  \pm 1 \\
  0 \\
  0 \\
\end{array}%
\right). \]
It is a straightforward calculation to find the solution for $y(s)$ and impose the boundary conditions. We find then that $A$ must satisfy the condition 
\[ \cos \sqrt{A^2 + \frac{w^2}{(1-r)^2}}= \pm 1 \] 
which implies that
\[ A=A(m,n)= \sqrt{(m \pi)^2 -  \frac{w^2}{(1-r)^2}}= \pi \, \sqrt{m^2 - 
\frac{(n+1/2)^2}{(1-r)^2}}, \]
where $m \in \mathbb{N}$ is such that $m \geq \frac{(n+1/2)}
{(1-r)}$. The solution for the state evolution is given by
\begin{eqnarray*}
&& \left(%
\begin{array}{r}
  \bar{x}_1 \\
 e^{i\phi_{12}}\, \bar{x}_2 \\
 e^{i(\phi_{12}+\phi_{23})}\,  \bar{x}_3 \\
\end{array}%
\right)(s) =    \\
&& \left(%
\begin{array}{lcr}
\cos ((n+\frac{1}{2})\pi s) & 0 & \sin ((n+\frac{1}{2})\pi s) \\ 
0 & 1 & 0 \\
-\sin ((n+\frac{1}{2})\pi s) & 0 & \cos ((n+\frac{1}{2})\pi s) \\
\end{array}%
\right) \left(%
\begin{array}{r}
  \cos(m \pi s) \\
  -A(m,n) \frac{\sin(m \pi s)}{m \pi} \\
  -\frac{n+1/2}{m(1-r)} \sin(m \pi s) \\
\end{array}%
\right). 
\end{eqnarray*}
We see that the local minimizers for this \otp are parameterized in terms of two integers, $n \in \mathbb{N}$ and $m \in \mathbb{N}$ such that $m \geq \frac{(n+1/2)}{(1-r)}$. The cost of such a minimizer can be easily computed to be equal to
\[ J=2 \pi^{2}\, [m^2 - \frac{(n+1/2)^2}{1-r}]. \]
The values of $m$ and $n$ and the exact cost for the global minimum are uniquely specified by the value of $r$.

\bibliographystyle{ieeetr}
\bibliography{myreferences}

\begin{thebibliography}{10}

\bibitem{khabrogla01a}
N.~Khaneja, R.~Brockett, and S.~Glaser, ``Time optimal control in spin
  systems,'' {\em Physical Review A}, vol.~63, no.~032308, February 2001.

\bibitem{dal02}
D.~D'Alessandro, ``The {O}ptimal {C}ontrol {P}roblem on $so(4)$ and {I}ts
  {A}pplications to {Q}uantum {C}ontrol,'' {\em IEEE Transactions on Automatic
  Control}, vol.~47, no.~1, January 2002.

\bibitem{khareiluygla02}
N.~Khaneja, T.~Reiss, B.~Luy, and S.~Glaser, ``Optimal control of spin dynamics
  in the presence of relaxation,'' {\em arXiv:quant-ph/0208050}, 2002.
\newblock Preprint.

\bibitem{daldah01}
D.~D'Alessandro and M.~Dahleh, ``Optimal control of two-level quantum
  systems,'' {\em IEEE Transactions on Automatic Control}, vol.~45, no.~1, June
  2001.

\bibitem{boschagau02}
U.~Boscain, T.~Chambrion, and J.~Gauthier, ``On the $k + p$ problem for a
  three-level quantum system: {O}ptimality implies resonance,'' {\em
  arXiv:math.OC/0204233}, April 2002.
\newblock Preprint.

\bibitem{boschagauguejau02}
U.~Boscain, T.~Chambrion, J.~Gauthier, S.~Gu\'{e}rin, and H.~Jauslin, ``Optimal
  control in laser-induced population transfer for two- and three-level quantum
  systems,'' {\em Journal of Mathematical Physics}, vol.~43, p.~2107, 2002.

\bibitem{sheshirab93}
L.~Shen, S.~Shi, and H.~Rabitz, ``Control of coherent wave functions: {A}
  linearized molecular dynamics view,'' {\em Journal of Physical Chemistry},
  no.~97, p.~8874, 1993.

\bibitem{jur01}
V.~Jurdjevic, ``Hamiltonian point of view of non-{E}uclidean geometry and
  elliptic functions,'' {\em Systems \& {C}ontrol {L}etters}, no.~43,
  pp.~25--41, 2001.

\bibitem{peidahrab87}
A.~Peirce, M.~Dahleh, and H.~Rabitz, ``Optimal control of quantum-mechanical
  systems: {E}xistence, numerical approximation, and applications,'' {\em
  Physical Review A}, vol.~37, no.~12, p.~4950, 1987.

\bibitem{zhurab98}
W.~Zhu and H.~Rabitz, ``A rapid monotonically convergent iteration algorithm
  for quantum optimal control over the expectation value of a positive definite
  operator,'' {\em Journal of Chemical Physics}, vol.~109, no.~2, p.~385, 1998.

\bibitem{gribam02}
S.~Grivopoulos and B.~Bamieh, ``Iterative algorithms for optimal control of
  quantum systems,'' in {\em Proceedings of the 41st {IEEE} {C}onference on
  {D}ecision and {C}ontrol}, December 2002.

\bibitem{ramsaldahrabpei95}
V.~Ramakrishna, M.~Salapaka, M.~Dahleh, H.~Rabitz, and A.~Peirce,
  ``Controllability of molecular systems,'' {\em Physical Review A}, vol.~51,
  no.~2, p.~960, 1995.

\bibitem{alt02a}
C.~Altafini, ``Controllability of quantum mechanical systems by root space
  decomposition of su({N}),'' {\em Journal of Mathematical Physics}, vol.~43,
  no.~5, p.~2051, 2002.

\bibitem{turrab01}
G.~Turinici and H.~Rabitz, ``Quantum wavefunction controllability,'' {\em
  Journal of Chemical Physics}, vol.~267, p.~1, 2001.

\bibitem{jursus72}
V.~Jurdjevic and H.~Sussmann, ``Control {S}ystems on {L}ie {G}roups,'' {\em
  Journal of Differential Equations}, no.~12, p.~313, 1972.

\bibitem{bro72}
R.~Brockett, ``System {T}heory on {G}roup {M}anifolds and {C}oset {S}paces,''
  {\em SIAM Journal of Control}, vol.~10, no.~2, p.~265, 1972.

\bibitem{bro73}
R.~Brockett, ``Lie {T}heory and {C}ontrol {S}ystems defined on {S}pheres,''
  {\em SIAM Journal of Applied Mathematics}, vol.~25, no.~2, p.~213, 1973.

\bibitem{jur97}
V.~Jurdjevic, {\em Geometric control theory}.
\newblock Cambridge University Press, 1997.

\bibitem{sagwhi77}
A.~P. Sage and C.~C. White, III, {\em Optimum {S}ystems {C}ontrol}.
\newblock Prentice-Hall, second~ed., 1977.

\bibitem{boscha03}
U.~Boscain and G.~Charlot, ``Resonance of minimizers for n-level quantum
  systems with an arbitrary cost,'' {\em arXiv:quant-ph/0308103}, August 2003.
\newblock Preprint.

\bibitem{kha96}
H.~Khalil, {\em Nonlinear {S}ystems}.
\newblock Prentice {H}all, second~ed., 1996.

\bibitem{arn73}
V.~Arnol'd, {\em Ordinary {D}ifferential {E}quations}.
\newblock Cambridge, MIT Press, 1973.

\end{thebibliography}
\end{document}